%% file: main.tex
\documentclass[sigconf]{acmart}
\AtBeginDocument{%
  }

\copyrightyear{2025} 
\acmYear{2025} 
\setcopyright{acmlicensed}\acmConference[CHI '25]{CHI Conference on Human Factors in Computing Systems}{April 26-May 1, 2025}{Yokohama, Japan}
\acmBooktitle{CHI Conference on Human Factors in Computing Systems (CHI '25), April 26-May 1, 2025, Yokohama, Japan}
\acmDOI{10.1145/3706598.3714273}
\acmISBN{979-8-4007-1394-1/25/04}

\usepackage{subcaption}
\usepackage{graphicx}
\usepackage{multicol}
\usepackage{enumitem}
\usepackage[title]{appendix}
\usepackage{url}
\usepackage{algorithm}
\usepackage{algpseudocode}

\newcommand*{\SuperScriptSameStyle}[1]{%
  \ensuremath{%
    \mathchoice
      {{}^{\displaystyle #1}}%
      {{}^{\textstyle #1}}%
      {{}^{\scriptstyle #1}}%
      {{}^{\scriptscriptstyle #1}}%
  }%
}
\newcommand*{\oneS}{\SuperScriptSameStyle{*}}
\newcommand*{\twoS}{\SuperScriptSameStyle{**}}
\newcommand*{\threeS}{\SuperScriptSameStyle{*{*}*}}

\begin{document}

\title[VibraForge]{VibraForge: A Scalable Prototyping Toolkit For Creating Spatialized Vibrotactile Feedback Systems}

\author{Bingjian Huang}
\email{bingjian20@dgp.toronto.edu}
\affiliation{%
  \institution{Dynamic Graphics Project Lab, University of Toronto}
  \city{Toronto}
  \state{Ontario}
  \country{Canada}
}

\author{Siyi Ren}
\email{siyi.ren@mail.utoronto.ca}
\affiliation{%
  \institution{Division of Engineering Science, University of Toronto}
  \city{Toronto}
  \state{Ontario}
  \country{Canada}
}

\author{Yuewen Luo}
\email{sofia.luo@mail.utoronto.ca}
\affiliation{%
  \institution{Division of Engineering Science, University of Toronto}
  \city{Toronto}
  \state{Ontario}
  \country{Canada}
}

\author{Qilong Cheng}
\email{qilong.cheng@mail.utoronto.ca}
\affiliation{%
  \institution{Mechanical Engineering, University of Toronto}
  \city{Toronto}
  \state{Ontario}
  \country{Canada}
}

\author{Hanfeng Cai}
\email{hanfeng.cai@mail.utoronto.ca}
\affiliation{%
  \institution{Department of Electrical and Computer Engineering, University of Toronto}
  \city{Toronto}
  \state{Ontario}
  \country{Canada}
}

\author{Yeqi Sang}
\email{abel.sang@mail.utoronto.ca}
\affiliation{%
  \institution{Department of Mechanical \& Industrial Engineering, University of Toronto}
  \city{Toronto}
  \state{Ontario}
  \country{Canada}
}

\author{Mauricio Sousa}
\email{mauricio@dgp.toronto.edu}
\affiliation{%
  \institution{Dynamic Graphics Project Lab, University of Toronto}
  \city{Toronto}
  \state{Ontario}
  \country{Canada}
}

\author{Paul H. Dietz}
\email{dietz@cs.toronto.edu}
\affiliation{%
  \institution{Dynamic Graphics Project Lab, University of Toronto}
  \city{Toronto}
  \state{Ontario}
  \country{Canada}
}

\author{Daniel Wigdor}
\email{daniel@dgp.toronto.edu}
\affiliation{%
  \institution{Dynamic Graphics Project Lab, University of Toronto}
  \city{Toronto}
  \state{Ontario}
  \country{Canada}
}

\renewcommand{\shortauthors}{Huang et al.}

\input{docs/0_abstract}


\begin{teaserfigure}
  \includegraphics[width=\textwidth]{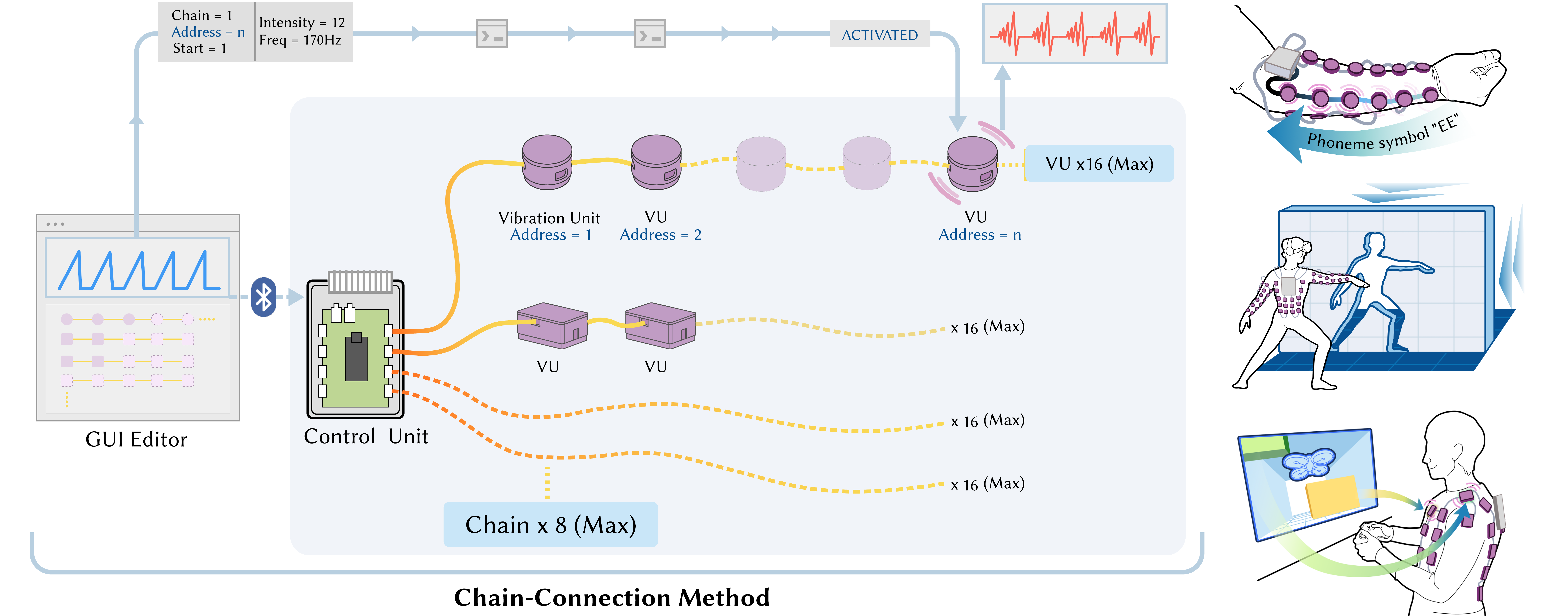}
  \caption{VibraForge is a scalable, open-source vibrotactile toolkit that supports the creation of spatialized vibrotactile feedback systems. Leveraging a chain-connection method, one control unit in the system can independently control up to 128 vibration units. Applications of the toolkit include phonemic tactile displays, VR fitness training, and drone teleoperation.}
  \Description{An overview of the VibraForge toolkit, which has control units and vibration units, uses a chain-connection topology, and supports the creation of spatialized vibrotactile feedback systems.}
  \label{fig:teaser}
\end{teaserfigure}

\maketitle

\input{docs/1_introduction}
\input{docs/2_relatedwork}
\input{docs/3_hardwaredesign}
\input{docs/4_technicalevaluation}
\input{docs/5_casestudyphonemic}

\input{docs/5_casestudyvr}
\input{docs/6_casestudydrone}
\input{docs/RR_usabilitystudy}
\input{docs/7_discussion}

\input{docs/8_futurework}
\input{docs/9_conclusion}

\begin{acks}
% Mengfei for sewing machine and advice.
% Jeb, Brenna for feedback.
% Michelle for paper editing.
% forreal graphics.
% GPT editing
We would like to express our deepest gratitude to Dr. Michelle Annett, whose assistance significantly enhanced the quality of our manuscript. We would also like to thank Jeb Thomas and Brenna Li for providing feedback on paper writing. We also thank Mengfei Liu, who kindly provided the sewing machine for tailoring the garments used in this work. Some figures in the manuscript were designed and improved by For Real Entertainment and Annie Squarecircle. ChatGPT was used in the writing process for language enhancement.
\end{acks}

\bibliographystyle{ACM-Reference-Format}
\bibliography{sample-base}

\end{document}

%% file: docs/0_abstract.tex
\begin{abstract}

% Problem
Spatialized vibrotactile feedback systems deliver tactile information by placing multiple vibrotactile actuators on the body. As increasing numbers of actuators are required to adequately convey information in complicated applications, haptic designers find it difficult to create such systems due to limited scalability of existing toolkits. 
% solution
We propose VibraForge, an open-source vibrotactile toolkit that supports up to 128 vibrotactile actuators. Each actuator is encapsulated within a self-contained vibration unit and driven by its own microcontroller. By leveraging a chain-connection method, each unit receives independent vibration commands from a control unit, with fine-grained control over intensity and frequency. We also designed a GUI Editor to expedite the authoring of spatial vibrotactile patterns.
% evaluation
Technical evaluation showed that vibration units reliably reproduced audio waveforms with low-latency and high-bandwidth data communication. Case studies of a phonemic tactile display, virtual reality fitness training, and drone teleoperation demonstrated the potential usage of VibraForge within different domains. A usability study with non-expert users highlighted the low technical barrier and customizability of the toolkit.

\end{abstract}

%% file: docs/1_introduction.tex
\section{Introduction}

% version 20240701

% background
Tactile feedback conveys sensory information to the human body through the sense of touch~\cite{weber1996eh}. Vibrotactile feedback, which produces tactile sensations via actuators vibrating on the skin, has gained much popularity because it is compact, affordable, and accessible, as summarized by Choi et al.~\cite{choi2013vibrotactile}. It has been widely adopted in various domains, such as virtual reality (VR) \cite{Wang2021Research,kaul2017haptichead,bHapticsWebsite}, robotics \cite{pacchierotti2024cutaneous,xia2022virtual,aggravi2018design}, rehabilitation \cite{bark2014effects,prabhu2020vibrosleeve,lindeman2006tactapack}, and accessibility \cite{witteveen2015vibrotactile,jung2022wireless,reed2018phonemic}.
%and immersive experience and multimedia entertainment \cite{kaul2017increasing, Oliveira2017Designing, han2023generating}.

% Quote from phonemic display: there is still a need for the development of wearable tactile devices as aids to communication. Such devices would have applications to a broad range of situations where input to the auditory and/or visual sense is absent or compromised, or when these sensory channels are overloaded in performing other tasks.

To provide adequate feedback to users during complex tasks, spatialized vibrotactile feedback systems have been proposed, such as a battle simulation suit by Lindeman et al.~\cite{lindeman2004towards} and directed movement sleeves by Louison et al.~\cite{louison2017spatialized}. These systems utilize actuator positions as an additional channel to enhance information transmission. For example, Wang et al. designed a facial mask with 36 eccentric rotating mass (ERM) actuators to study the perception of localization and orientation of objects in VR ~\cite{Wang2021Research}. The bHaptics TactSuit X40~\cite{bHapticsWebsite} utilized 40 ERMs on a vest to create localized feedback for immersive VR gaming experiences. The bHaptics TactSuit was also used by Xia et al. to convey hydrodynamic flow intensity during submarine teleoperation~\cite{xia2022virtual}. Jung et al. designed a wireless haptic interface with 36 ERMs to convey detailed fingertip touch sensations on the upper arm for amputees~\cite{jung2022wireless}. Sanchez et al. created OpenVNAVI, a navigation aid system with 128 ERMs to guide visually impaired people~\cite{OpenVNAVI}. A 64-actuator tactile jacket was also created by Lemmens et al. to enrich movie viewing experiences~\cite{lemmens2009body}. Similarly, West et al. created the Body:Suit:Score system with 60 ERMs to enhance musical experiences~\cite{west2019design}.

% problem
Despite the great potential of spatialized vibrotactile feedback systems, previous implementations of such systems revealed scalability and expressivity limitations. Systems that depend on direct connections between actuators and microcontroller units (MCUs) exhibit limited scalability, as the number of actuators is intrinsically restricted by the available GPIOs or PWM channels on a MCU~\cite{Wang2021Research,kaul2017haptichead,reed2018phonemic}. Expanding the number of actuators requires multiple MCUs, resulting in exponential increases in system complexity and communication overhead. For systems that use custom drivers and data bus protocols to indirectly drive actuators, multi-layer connections support more actuators but compromise vibration expressivity \cite{li2013design, lemmens2009body, west2019design}. These systems cannot render complex waveforms such as audio signals due to bandwidth limits, thereby limiting the types of actuators that can be used to simple actuators like ERMs. Moreover, while various toolkits have been developed to assist vibrotactile design, they typically employ direct or multi-layer connections and face the same limitations (e.g., TECHTILE~\cite{minamizawa2012techtile}, VITAKI~\cite{martinez2014vitaki}, and VHP~\cite{dementyev2021vhp}). 
% GUIs for vibrotactile design focus on editing waveforms for a single actuator, rather than authoring multi-actuator spatial patterns.

% solution
To overcome these limitations, this work proposes VibraForge, an open-source vibrotactile toolkit\footnote{The project resource can be found at: \url{https://github.com/huangbj16/VibraForge}} that leverages a scalable, modular design to enable the rapid prototyping of vibrotactile systems. VibraForge consists of control units and vibration units. Control units receive vibration control commands from a Bluetooth server and sending them to vibration units. Vibration units are self-contained modules with an actuator driven by their own microcontroller. They are connected to control units via a chain-connection method, where a single GPIO pin on a control unit can control up to sixteen vibration units via a custom UART protocol. To better support spatial pattern authoring, we also designed an accompanying GUI Editor that enables haptic designers to intuitively author multi-actuator tactile patterns. 

Technical evaluation of VibraForge showed that the toolkit has a high bandwidth (200 Hz), low latency (16 ms), and can reliably render complex audio signals in real time. Through three case studies, we demonstrate that VibraForge not only supports the efficient replication of existing research prototypes such as phonemic displays \cite{reed2018phonemic}, but also improves user performance in various applications such as virtual reality fitness gaming and robot teleoperation. In a usability study, ten participants with diverse technical backgrounds learned to use the toolkit in 15 minutes and completed design tasks and a free exploration activity. The findings from this study reaffirmed VibraForge's low technical barrier and high potential for customizability.

The contributions of this research are:
\begin{itemize}
    \item an open-source hardware toolkit that supports the design of spatialized vibrotactile feedback systems,
    \item a novel chain-connection method that addresses the gap between system scalability and feedback expressivity,
    \item a software GUI editor that enables designers to intuitively author spatial vibration patterns,
    \item technical evaluations of VibraForge, including latency, bandwidth, vibration characteristics, etc., and 
    \item case studies and a usability study demonstrating the usefulness and usability of the toolkit.
\end{itemize}

% paper structure
%1. Introduction
%2. Related work
%    vibrotactile prototyping toolkit
%    commercial devices? aka existing multi-actuator systems, and their limitations
%    challenge of scalability: tradeoff between expressivity and scalability
%    software GUI references
%    vibrotactile application domains
%    data protocols?
%3. Design
%    design principles
%    vibration unit
%        easy assembly 
%        flexible config
%        solderless, reliable
%        extend to more actuators
%    control unit
%        different sizes to fit different body mounting positions
%        bluetooth wireless, 
%    data communication
%        chain connection topology
%        custom UART protocol
%    GUI design
%4. Technical evaluation
%5.1 what about using VibraForge to replicate some former systems?
%5.2 Case studies
%    VR fitness
%        garment design,
%        vibrations on the sleeve
%    UAV teleoperation
%        vibration on the body
%        integration into UAV workflow
%6. Discussion
%7. Conclusion    

%% file: docs/2_relatedwork.tex
\begin{figure*}[!ht]
    \centering
    \includegraphics[width=\linewidth]{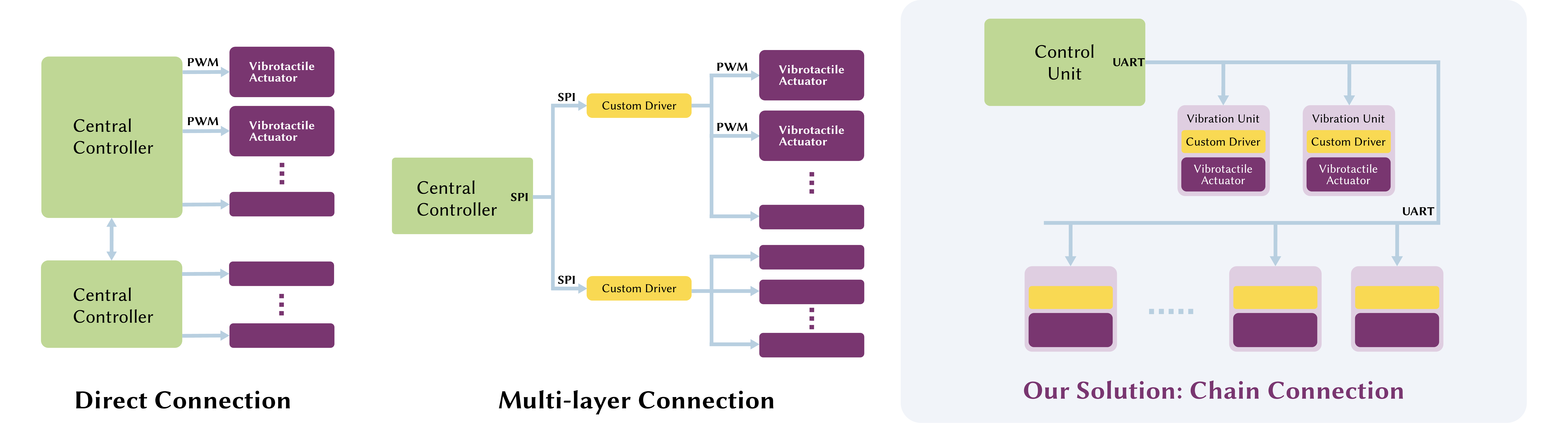}
    \caption{Depiction of the direct connection (left) and multi-layer connection (middle) methods commonly used in spatialized vibrotactile systems and our proposed solution (right), which uses a chain connection method.}
    \Description{This figure illustrates three different connection methods for vibrotactile actuators: direct connection using PWM, multi-layer connection with SPI and custom drivers, and a chain connection solution using UART between vibration units for scalability.}
    \label{fig:sec2_connection_topology}
\end{figure*}

\section{Related Work}

Herein, we provide an overview of existing vibrotactile toolkits, highlighting their strengths and limitations. Then, we summarize previous spatialized vibrotactile feedback systems and discuss the benefits and drawbacks of their design strategies. Lastly, we summarize existing software for vibration pattern design.

\subsection{Existing Vibrotactile Toolkits}

% define toolkit with Saul Greenberg's paper

Vibrotactile actuators such as eccentric rotation mass vibration motors (ERM), linear resonant actuators (LRA), voice coil actuators (VCA), or piezoelectric actuators (PIEZO) create mechanical vibrations via electromagnetic motions. Each actuator type and model has distinct voltage ratings, frequency responses, and vibration characteristics, making it challenging for designers to explore different configurations when creating wearable tactile devices. To assist their design process, various vibrotactile toolkits have been proposed. These toolkits abstract low-level technical details and offer intuitive design interfaces, so designers can concentrate on creating vibration effects without being overwhelmed by the complexities of electronic, mechanical, and software design.

% An early example from industry would be the Texas Instrument DRV2605L driver [XXX], which was has PWM input, and offers closed-loop control and an integrated waveform library, making it a popular choice for initial prototyping. 

Vibrotactile prototyping toolkits can be categorized into those for education purposes and those for design purposes. Toolkits for education purposes often have simple compact widgets and limited output channels so they can teach novice users about the basics of haptic feedback and the design of simple vibration effects. For example, the TECHTILE toolkit was composed of a microphone, a signal amplifier, and two actuators \cite{minamizawa2012techtile}. Audio signals were captured by the microphone, amplified through the amplifier, and delivered to the actuators. The Stereohaptics toolkit used a similar principle, enabling users to create, record, and playback simple haptic waveforms, making haptic design accessible through familiar audio tools \cite{israr2016stereohaptics}. 

Toolkits for design purposes often have more sophisticated hardware and complex functionality that enable researchers and designers to explore different configurations of actuators and vibration patterns. For example, VITAKI was an ERM-based toolkit targeting virtual reality and video game applications \cite{martinez2014vitaki}. It used shift registers and PWM drivers to control up to sixteen ERMs (i.e., the number of channels available on the PWM drivers). TactJam supported up to eight actuators and was intended for collaborative design \cite{wittchen2022tactjam}. With Syntacts, designers could design vibration effects in a GUI Editor and then send them through an audio amplifier to actuators for testing \cite{pezent2020syntacts}. The Syntacts audio amplifier, however, only supported eight channels and was too bulky to wear on the body. To enable the creation of wearable devices, the VHP toolkit used a custom designed PCB and flexible PCB connectors so devices could fit to different body shapes \cite{dementyev2021vhp}. The Nordic nRF52840 microcontrollers used in the toolkit had three PWM modules that generated 12 channels of PWM signals to drive the actuators.

While existing toolkits provide rich features to support vibrotactile design, they are limited by their scalability, particularly in the number of actuators they can support. This is because the actuators are directly controlled via GPIO pins or PWM channels on MCUs, with most offering a maximum of 16 channels. As vibrotactile feedback continues to be used for more complicated tasks and applications, these toolkits cannot meet the demands of emerging systems. Therefore, when designing VibraForge, we drew insights from previous spatialized vibrotactile feedback systems and created a new connection method to overcome these scalability limitations.

% prior work

\subsection{Spatialized Vibrotactile System Design}

Two connection methods are often used when building spatialized vibrotactile systems (Figure~\ref{fig:sec2_connection_topology}). Some have tried maintaining a "direct connection" to each actuator by adding more microcontrollers to a system, thus increasing the number of PWM channels. For example, Wang et al. used PWM outputs from four Arduino Mega2560 boards to drive their facial display \cite{Wang2021Research}. Similarly, Kaul et al. used four Arduino Nanos to drive 17 actuators mounted on their HapticHead display \cite{kaul2017haptichead}. Reed et al. used three custom-built audio amplifier boards to send individual waveforms to 24 actuators on a sleeve to deliver speech phonemic cues to deaf users \cite{reed2018phonemic}. Although complex waveforms could be sent to the actuators via direct connections, communication overhead and system complexity increased significantly when the system was scaled up.

%cite: A Research on Sensing Localization and Orientation of Objects in VR with Facial Vibrotactile Display
% HapticHead
% Measuring relative vibrotactile spatial acuity: effects of tactor type, anchor points and tactile anisotropy
% bHaptics suit
% Towards Full-Body Haptic Feedback:
% Virtual Training via Vibrotactile Arrays
% Green technology and wearable haptic feedback display with 5×12 arrays of vibrotactile actuators
% Providing Directional Information with Tactile Torso Displays

% Such design can also be seen in the industry, e.g., the bHaptics TactSuit \cite{bHapticsWebsite} links all 40 actuators on the body directly to the central unit on the back, resulting in an intricate wiring structure. 

Others have explored using "multi-layer connections" that employ custom driver boards to interface between the microcontrollers and actuators \cite{OpenVNAVI, li2013design, lemmens2009body, west2019design}. On the input side, these driver boards received vibration commands from a main controller board via data bus protocols such as I2C \cite{OpenVNAVI} or SPI \cite{lemmens2009body,west2019design}. On the output side, they converted vibration commands to PWM signals and drove four \cite{lemmens2009body}, six \cite{west2019design}, eight \cite{OpenVNAVI}, or sixteen \cite{li2013design} actuators. By leveraging a multi-layer design, these systems were more scalable than previous solutions, however, the indirect control afforded by the data bus protocols made it impossible to process complex waveforms. This limited the expressivity of the actuators so only simple actuators like ERMs were used in these systems.

There is thus a trade-off between system scalability and actuator expressivity: a system that supports fine-grained control of individual actuators is difficult to scale, and vice versa. In this work, we propose a chain-connection method (Figure \ref{fig:sec2_connection_topology}), where actuators are encapsulated into self-contained vibration units and driven by their own PCB drivers. Chains of vibration units are connected to a central MCU. Each PCB driver receives commands from the central MCU and generates PWM signals for its own actuator, thus supporting scalability. The high bandwidth and low latency control of vibration characteristics enables actuators to render complex waveforms, thus enhancing expressivity.

\subsection{Software for Vibration Pattern Design}
Vibration pattern design is crucial in spatialized vibrotactile feedback systems. For a comprehensive review of previous design software and their public availability, please see \cite{terenti2023vireo}. Here, we discuss the common interfaces used to design single-actuator waveforms and multi-actuator patterns.

% Figure here showing the different design methods

For single-actuator design, designers often start with standard waveforms (e.g., sine, square, etc), multiply them to generate new waveforms, and modify them using envelopes, as illustrated in Syntacts~\cite{pezent2020syntacts}. While this method provides authoring flexibility, it requires that designers have a deep understanding of how waveforms can be composed through combinations of standard signals. This makes it difficult for novice designers to use. Others have created editors that enable the direct manipulation of waveforms. With these editors, designers have access to two parallel windows to edit vibration amplitude and frequency. For example, in the Macaron Editor, designers could directly move keyframe dots to modify a waveform, making the design process more intuitive \cite{schneider2016studying}. This method was later adopted by several industrial software tools, including Meta Haptics Studio~\cite{oculus2023haptics}. 

For multi-actuator authoring, the design focus shifts from waveform editing to pattern design. Designers compose vibration paths across multiple actuators on a canvas and the system converts the drawn paths to a series of vibration commands that are executed over time. For example, Schneider et al. utilized phantom sensation illusions to interpolate points between actuators to produce more continuous vibration sensations in Mango \cite{schneider2015tactile}.

VibraForge's GUI Editor combines single-actuator waveform design with a multi-actuator canvas to support the efficient design for spatialized systems. It provides a library of standard waveforms and envelopes for composing complex waveforms. It is also compatible with existing tools such as Syntacts \cite{pezent2020syntacts} and VIREO \cite{terenti2023vireo} to support the reuse of workflows and prior knowledge. The canvas shows a virtual layout of actuators that replicates the physical chain connections so designers can reference it while designing spatial patterns.

%% file: docs/3_hardwaredesign.tex
\section{VibraForge Design}

\label{sec3:hardware_design}

% version 20240705
As highlighted in prior work, a trade-off exists between actuator  expressivity and system scalability. To address this, we developed a novel chain-connection method with custom vibration units and control units. This section provides an overview of the VibraForge toolkit, covering its design requirements, components, chain-connection method, signal segmentation algorithm, and GUI Editor.

\subsection{Design Requirements}

% \begin{enumerate}
%     \item Modularity/extensibility
%     \begin{enumerate}[label=\alph*.]
%         \item Plug-and-play
%         \item Easily extend to other actuators
%     \end{enumerate}
%     \item Robustness (won’t easily break, reliable)
%     \item Scalability
%     \begin{enumerate}[label=\alph*.]
%         \item Support large number of actuators
%         \item With end-to-end unnoticeable latency (human perception 20ms), ignorable voltage drop
%         \item Support multi-actuator vibrations
%     \end{enumerate}
%     \item Expressivity/expressiveness
%     \item Flexibility/reconfigurability
%     \item Portability/wearable, wireless
%     \item Usability
%     \item Compatibility (don’t reinvent wheels)
% \end{enumerate}

% DRs for two types: for wearable device, and for spatial vibrotactile
% Wearable asks for 5,6, flex design, portable design, easy attachment, repair and replacement;
% Spatial vibrotactile asks for 2,3,4,7,8, robustness, scalability, fine control, easy pattern design (with compactibility).

VibraForge's design requirements were drawn from design challenges discussed in prior work on spatial vibrotactile and wearable design, or identified by the authors based on prior prototyping experiences.

\begin{itemize}
    \item \textbf{DR1: Scalability}. The toolkit should support large quantities of actuators such as ERM, LRA, and VCA.
   
    \item \textbf{DR2: Fine-Grained Control}. The toolkit should provide fine-grained control of individual actuator vibration characteristics such as intensity and frequency to support expressivity. Regardless of the number of actuators employed in the system, the control mechanism should have a high bandwidth and low latency.
    
    \item \textbf{DR3: Usability}. The toolkit should provide an intuitive design interface for authoring vibrotactile spatial patterns that leverages designers' prior design knowledge and is compatible with existing vibrotactile waveform editors such as Syntacts \cite{pezent2020syntacts}, VIREO \cite{terenti2023vireo}, and Meta Haptics Studio \cite{oculus2023haptics}.

    \item \textbf{DR4: Flexibility}. The toolkit should enable designers to modify settings, reprogram behaviors, and physically rearrange components without difficulty, derived from design challenges faced by West et al. \cite{west2019design}. 

    \item \textbf{DR5: Portability}. Devices built with the toolkit should be powered by batteries, communicate with external devices via wireless protocols, and use lightweight components to ensure a comfortable wearing experience.
\end{itemize}

\subsection{Vibration Units}

% \begin{enumerate}
%     \item Self-contained, actuator, PCB, enclosure (R1)
%     \item No soldering (R2)
%     \item sPWM (R4)
%     \item Size of a coin, lightweight, unobtrusive, comfortable (R7)
%     \item Easily copy paste to other actuators (R1)
%     \item Preprogrammed, no address needed
%     \item Only 4 wires (R2)
%     \item Ring design, so it can be attached to clothes (R5)
% \end{enumerate}

\begin{figure}[h]
    \centering
    \includegraphics[width=\linewidth]{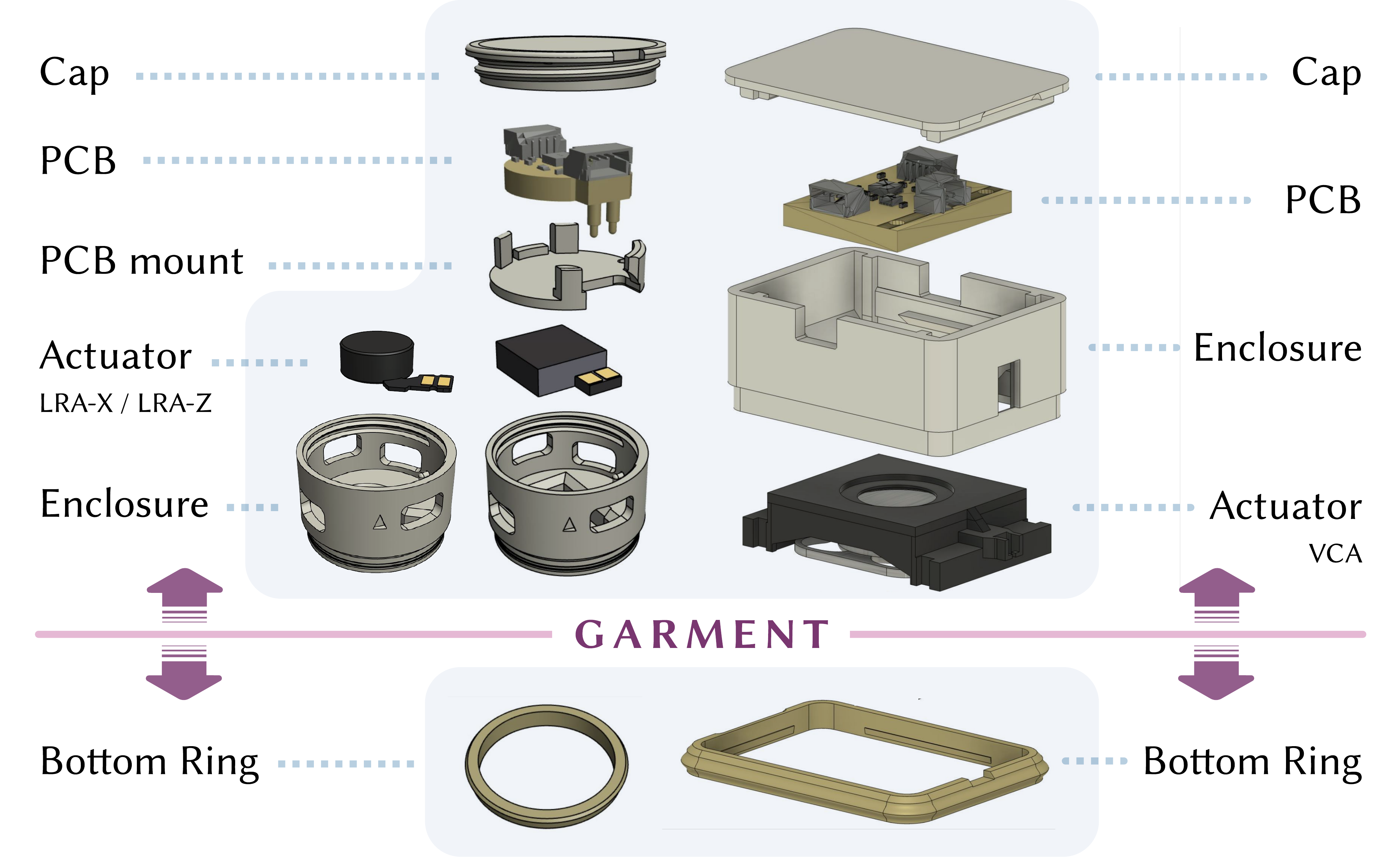}
    \caption{(Left) Exploded views of the small vibration unit for X-axis LRAs, including the cap, the PCB, the PCB mount, the enclosure, and the bottom ring. Z-axis LRAs shared most components except for the enclosure. (Right) Exploded view of the large vibration unit for VCAs with a similar structure. Note that the bottom ring was placed underneath the garment to attach the unit.}
    \Description{An exploded view of the LRA and VCA vibration units, showing the actuator, PCB, and 3D printed parts.}
    \label{fig:ch3_exploded_view}
\end{figure}

The vibrations units were self-contained modules comprised of a vibrotactile actuator, a custom designed PCB, and 3D printed components (Figure \ref{fig:ch3_exploded_view}). Three versions were designed to support various actuators (DR1).

\subsubsection{Vibrotactile Actuators}

The vibration unit used a square-shaped X-axis LRA (9.5 mm x 9.5 mm, Model L959535, Jinlong Machinery and Electronics \cite{JinlongWebsite}), resonating at 170 Hz with a maximum current of 145 mArms and max acceleration of 0.65±0.12 Grms at 0.9 Vrms. Two additional actuators included a cylindrical Z-axis LRA (8 mm diameter, 3.2 mm height, Model G0832, Jinlong \cite{JinlongWebsite}) resonating at 235 Hz with 0.50±0.08 Grms of acceleration and a maximum current of 90 mArms at 1.8 Vrms, and a VCA (32 mm x 22 mm, Model HD-VA3222, PUI Audio Inc. \cite{PUIAudio}) with a 80 - 500 Hz response range and 2.52 Gp-p acceleration at 133 Hz, 1.5 Vrms. These actuators covered various vibration directions and intensities to meet the tactile perception needs of different body parts (i.e., spatial acuity \cite{elsayed2020vibromap} and perceived intensity \cite{verrillo1969sensation}). Since they share the same PCB schematics and similar enclosures, we will focus on the X-axis LRA vibration unit in the following sections, with the other units shown in Figure \ref{fig:ch3_exploded_view}.

\subsubsection{Printed Circuit Board Design}

\begin{figure}[h]
    \centering
   \includegraphics[width=\linewidth]{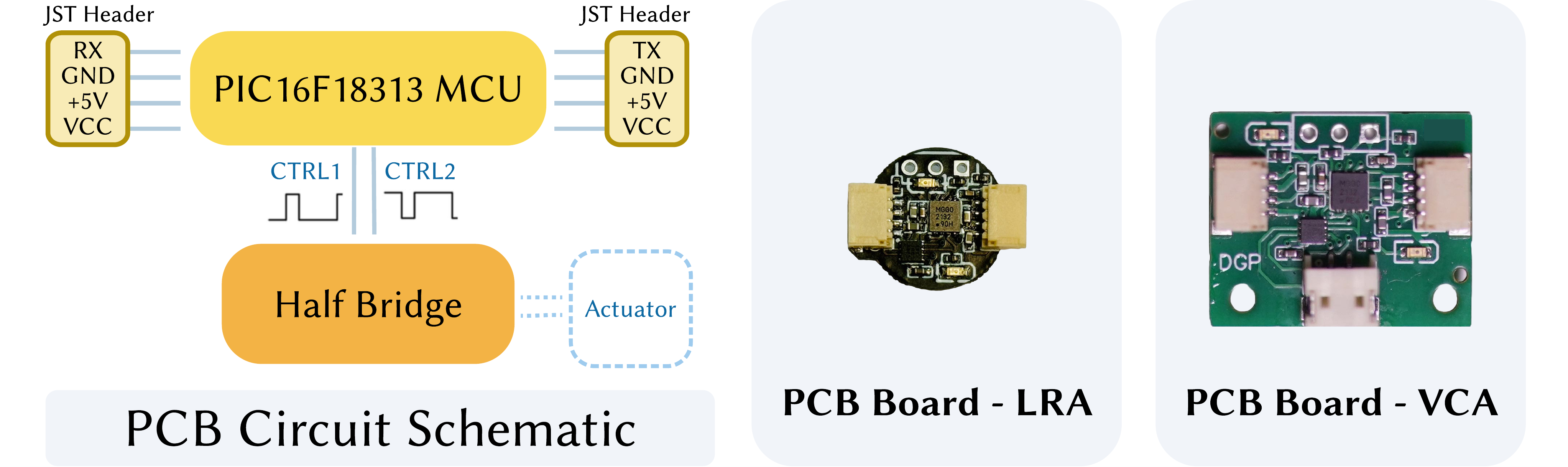}
    \caption{The circuit schematics and PCB board designs of the vibration units.}
    \Description{Circuit schematics of a unit driver that generated two opposite waveforms based on the received control signals to modulate the actuator.}
    \label{fig:ch3_pcb}
\end{figure}

Each vibration unit had a PCB driver board to receive UART commands from a control unit and provide fine-grained control over vibration characteristics (DR2). The PCB had a PIC16F18313 MCU, a DRV8837 H-bridge driver, two 4-pin JST-SH headers, and two pogo pins for actuator connections (Figure \ref{fig:ch3_pcb}). Upon receiving a UART command from the RX line of the JST connector, the MCU extracted vibration characteristics from the command, generated waveform using an internal PWM generator, and sent it to the actuator through the H-bridge driver. The communication protocol is detailed in Section \ref{DataCommunicationOnChains}. 

\subsubsection{3D Printed Enclosures}

The 3D printed enclosures were designed to enclose actuators and PCBs and allow easy attachment to garments (DR4). Compact and lightweight like a coin, they ensured a comfortable wearing experience (DR5). The enclosure included a cap with a screw-nut system to stabilize the PCB, a mounting plate for PCB protection and alignment of the pogo pins, a protective housing with two JST connector holes and heat dissipation vents, and a bottom ring that maintained direct skin contact without sewing (DR4) (Figure \ref{fig:ch3_exploded_view}). The same design strategy was also applied to design enclosures for other actuators.

\subsection{Control Unit}
\label{ControlUnit}
% \begin{enumerate}
%     \item Wireless/BLE comm (R6)
%     \item Different sizes, number of ports. Body attachment (R3, R5)
%     \item Separate power supply for MCUs and actuators (R2)
% \end{enumerate}

\begin{figure}[h]
\includegraphics[width=\linewidth]{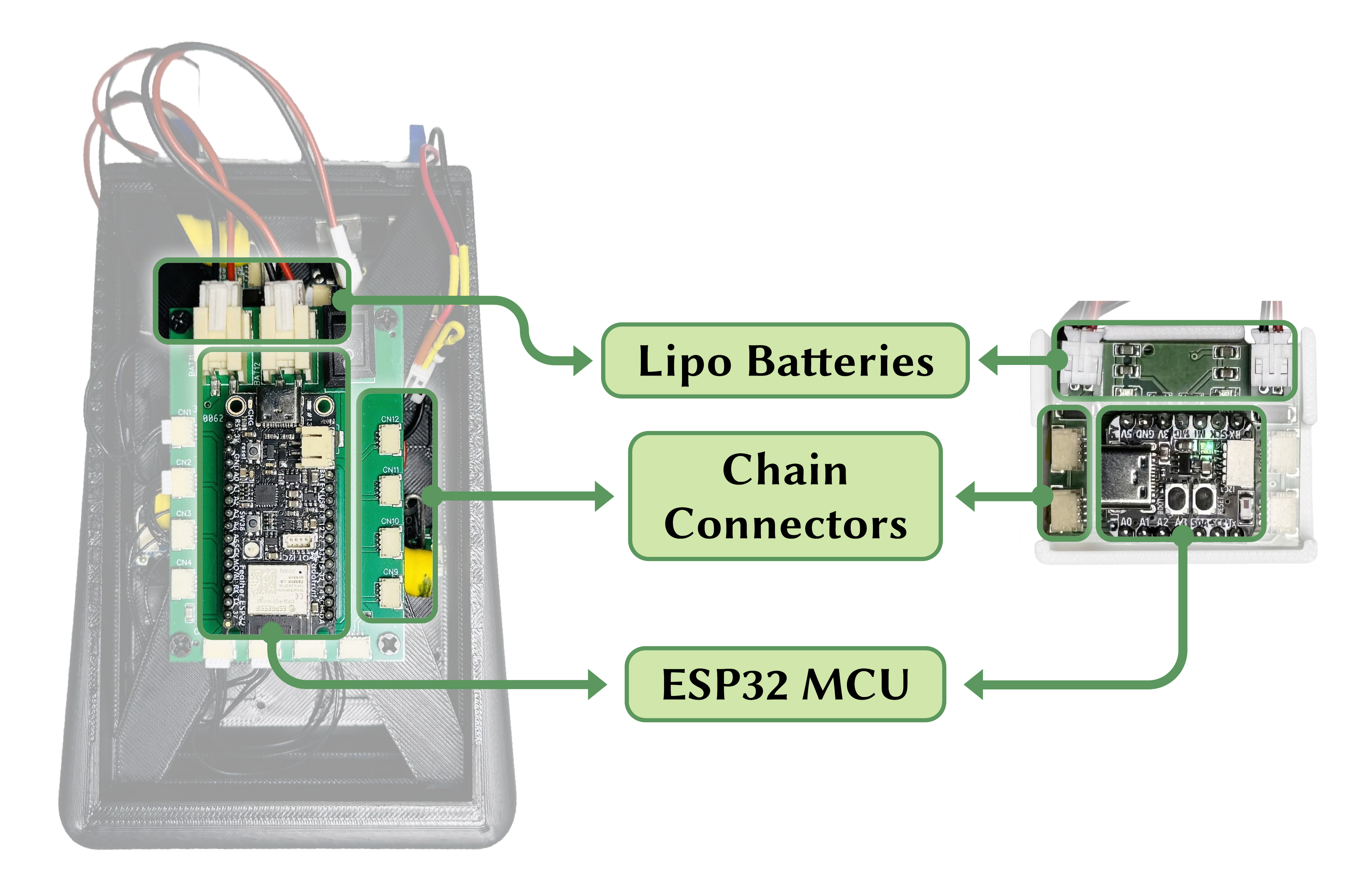}
\caption{The control unit, which contained an ESP32 MCU, an extension board, chain connectors, and LiPo batteries.}
\Description{This figure shows an ESP32 board mounted on the extension board in the middle of the control unit, and the 8 chain connection ports extending out from the control unit.}
\label{fig:ch3_control_unit}
\end{figure}

The portable control unit (DR5) featured a PCB extension board with ESP32 MCU, two LiPo batteries, and a 3D-printed enclosure (Figure \ref{fig:ch3_control_unit}). Multiple versions were designed to accommodate different project requirements, varying in the number of supported chains, the ESP32 model, and battery capacity. 

We designed a large control unit supporting up to 8 chains, with 16 units per chain (128 total). Powered by two 3.7 V 8200 mAh LiPo batteries, the unit used a linear regulator for the MCUs and a buck regulator for the actuators, ensuring 5 V stable voltage supplies for both. The extension board had eight JST connectors for vibration units, each linked to a GPIO pin on an Adafruit ESP32 Feather V2 MCU \footnote{https://learn.adafruit.com/adafruit-esp32-feather-v2}. The MCU converted Bluetooth commands to UART commands and sent to corresponding vibration units. 
For devices requiring fewer units, we also created a more compact control unit with two 3.7 V 500 mAh LiPo batteries and an Adafruit QT Py ESP32-S3 MCU \footnote{https://learn.adafruit.com/adafruit-qt-py-esp32-s3}, supporting four chain connections.
% https://learn.adafruit.com/adafruit-qt-py-esp32-s3/pinouts

\subsection{Chain-Connection Method and Data Communication}
\begin{figure}[h]
\includegraphics[width=\linewidth]{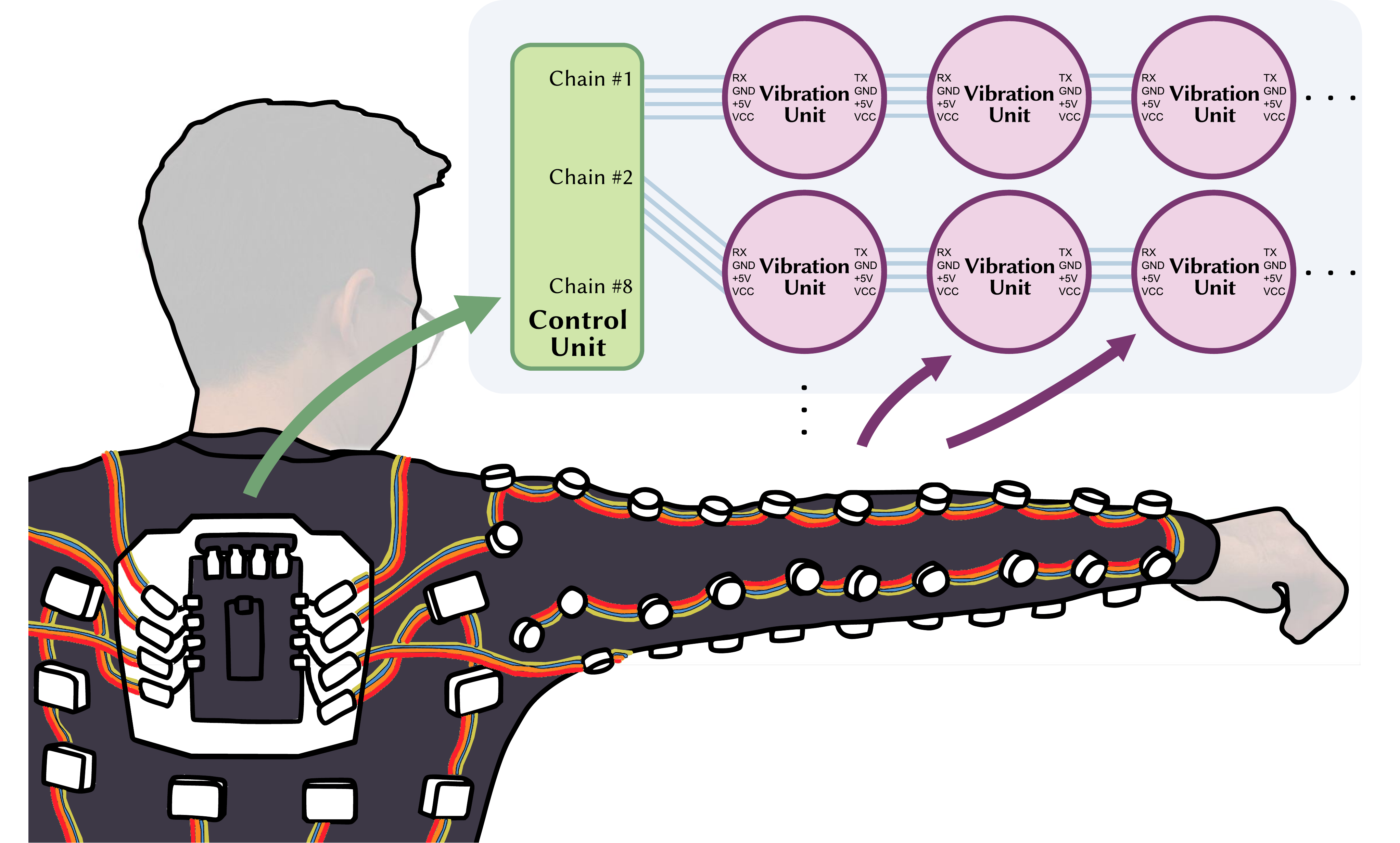}
\caption{The multiple chain-connection method used within VibraForge, depicting the connections required for a given chain.}
\label{fig:ch3_chain_connection}
\Description{This figure shows where the control unit, the power converters, and unit drivers are located to illustrate the multiple chain-connected method.}
\end{figure}

VibraForge differs from existing toolkits by using a chain-connection method (Figure \ref{fig:ch3_chain_connection}), allowing a single control unit to manage over 100 vibration units (DR1) and simultaneously control multiple actuators' vibration characteristics (DR2).

\label{DataCommunicationOnChains}

Vibration units are connected in chains to receive commands from the ESP32 GPIO pins. We considered two data transmission options (a) data bus protocols (e.g., CAN or I2C) would guarantee minimum latency via broadcasting commands, but require programming different addresses for each unit; (b) serial protocols (e.g., UART) did not require unique address, but passing commands along the chain would increase latency. We decided to use UART because it enabled plug-and-play control without extra programming (DR4) and latency could be reduced to imperceptible levels via high-speed transmission.

\begin{figure}[h]
\includegraphics[width=0.9\linewidth]{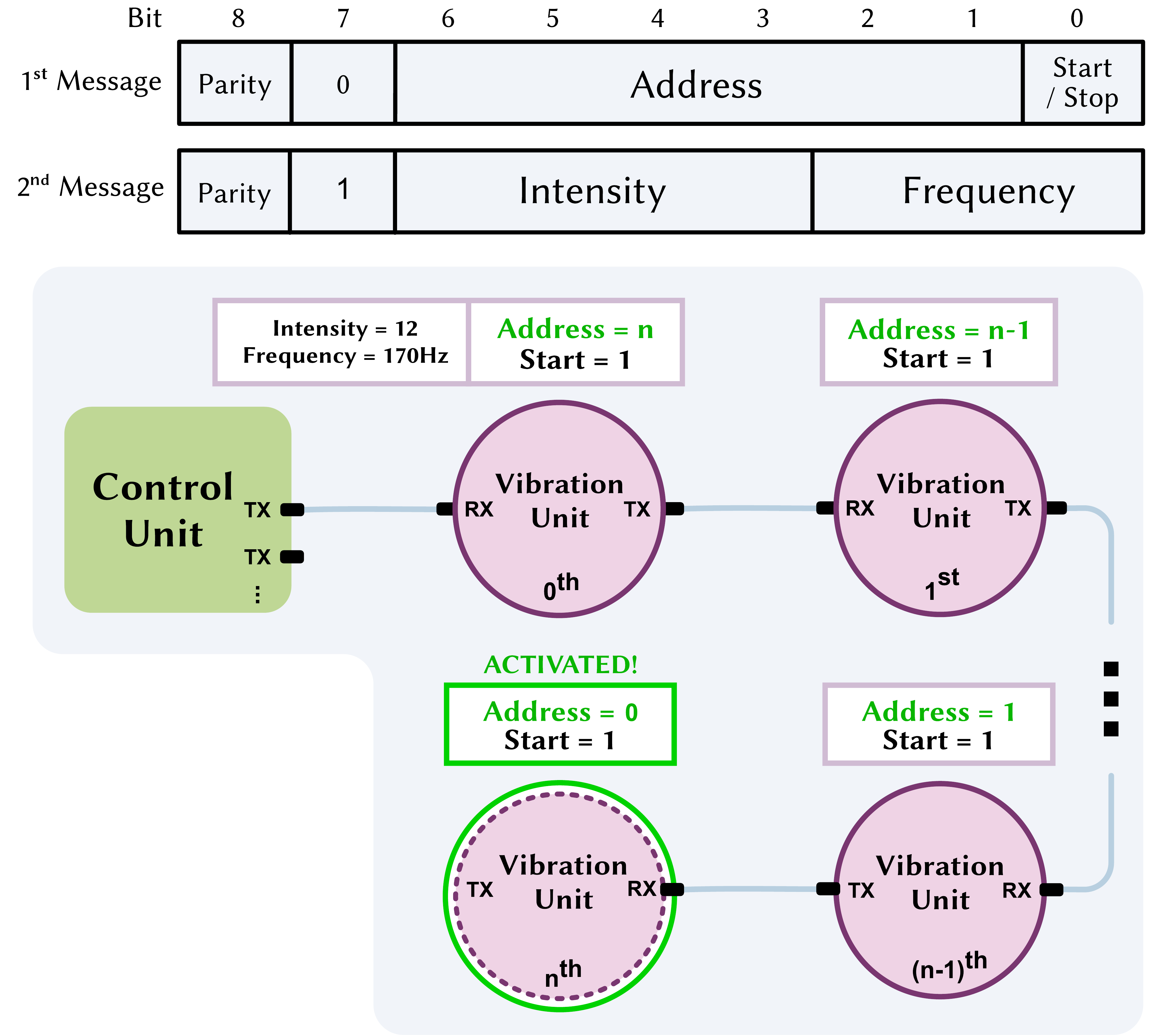}
\caption{The two-byte message bits definition and an example of how vibration commands were transmitted along a chain using the "virtual address" system.}
\Description{The figure shows the two-byte protocol that was developed to control each vibration unit. In both bytes, a 9-bit UART with parity-bit check was used to ensure message integrity. The address and start/stop bit were in the first message, and operational commands including the duty cycle, frequency, and waveform, were in the second message.}
\label{fig:ch3_protocol}
\end{figure}

Vibration commands were transmitted via UART at 115.2 kHz with parity checks (Figure \ref{fig:ch3_protocol}). Each command included two bytes: the first byte encoded the address and start/stop status, and the second byte specified vibration characteristics. Upon receiving the first byte, a unit checked the address, decremented it, and forwarded the command to the next unit. This repeated until the address was zero, thus signaling the target unit had been reached. The targe unit acted based on the start/stop bit: '0' (STOP) disabled the actuator, while '1' (START) activated it. The second byte detailed the vibration intensity  (16 levels from 0\% to 100\% ) and frequency (i.e., 8 levels following Weber's factor for frequency differences \cite{pongrac2006vibrotactile}, i.e., 123, 145, 170, 200, 235, 275, 322, and 384 Hz).

\subsection{Signal Segmentation Algorithm}

\begin{figure}[h]
    \centering
    \includegraphics[width=\linewidth]{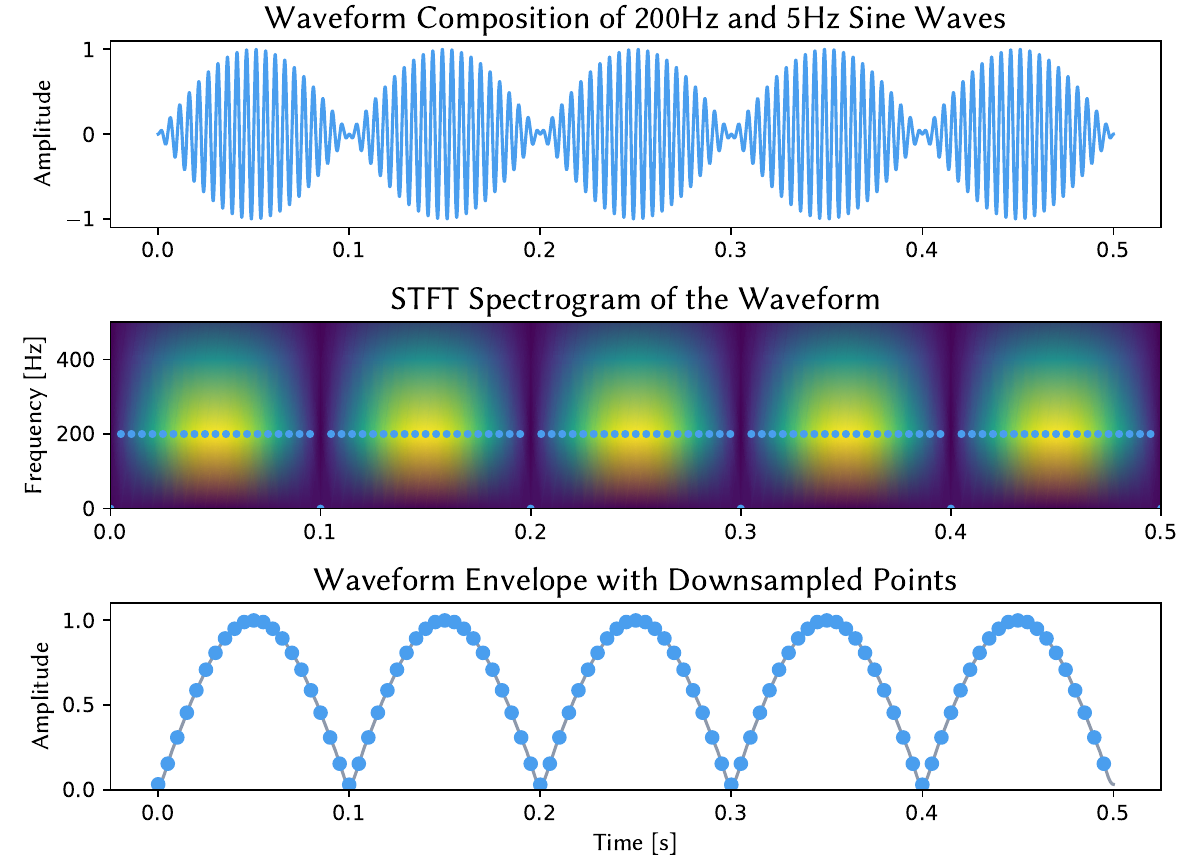}
    \caption{The signal segmentation workflow, showing how low frequency and high frequency components were extracted from a complex waveform.}
    \Description{The left figure shows multiplication waveform of 200Hz and 5Hz sine waves. The middle one shows STFT spectrogram with the highest energy surrounds 200Hz. The right one shows the envelope of the waveform.}
    \label{fig:sec3_signal_segmentation}
\end{figure}

To render complex waveforms like audio signals (DR2), we implemented a signal segmentation algorithm to convert high-frequency input signals to low-frequency serial commands (Figure \ref{fig:sec3_signal_segmentation}). For example, a 44100 Hz audio signal with 200 Hz and 5 Hz sine waves would be processed using a short-time Fourier transform (STFT) to extract high-frequency components (main frequency) and a Hilbert transform for low-frequency components (envelope). Both components were then downsampled to 200 Hz and sent to vibration units. Components above 100 Hz would be transmitted via frequencies, and those below 100 Hz as intensities. Technical evaluation (Section \ref{section4:technical_evaluation}) confirmed the algorithm 's reliability in rendering complex waveforms. 

\subsection{GUI Editor for Designing Spatialized Patterns}

% \begin{itemize}
%     \item provide simple sketching functions (R7)
%     \item enable import file from Syntacts, Interhaptics, MetaStudio, VIREO (R8)
% \end{itemize}

We designed a GUI Editor (Figure \ref{fig:sec3_gui_layout}) to support the creation of spatialized vibration patterns (DR3). The editor combined the waveform editing from single-actuator design software with the canvas design used for multi-actuator design. Its key components were:

\begin{figure}[h]
    \centering
    \includegraphics[width=\linewidth]{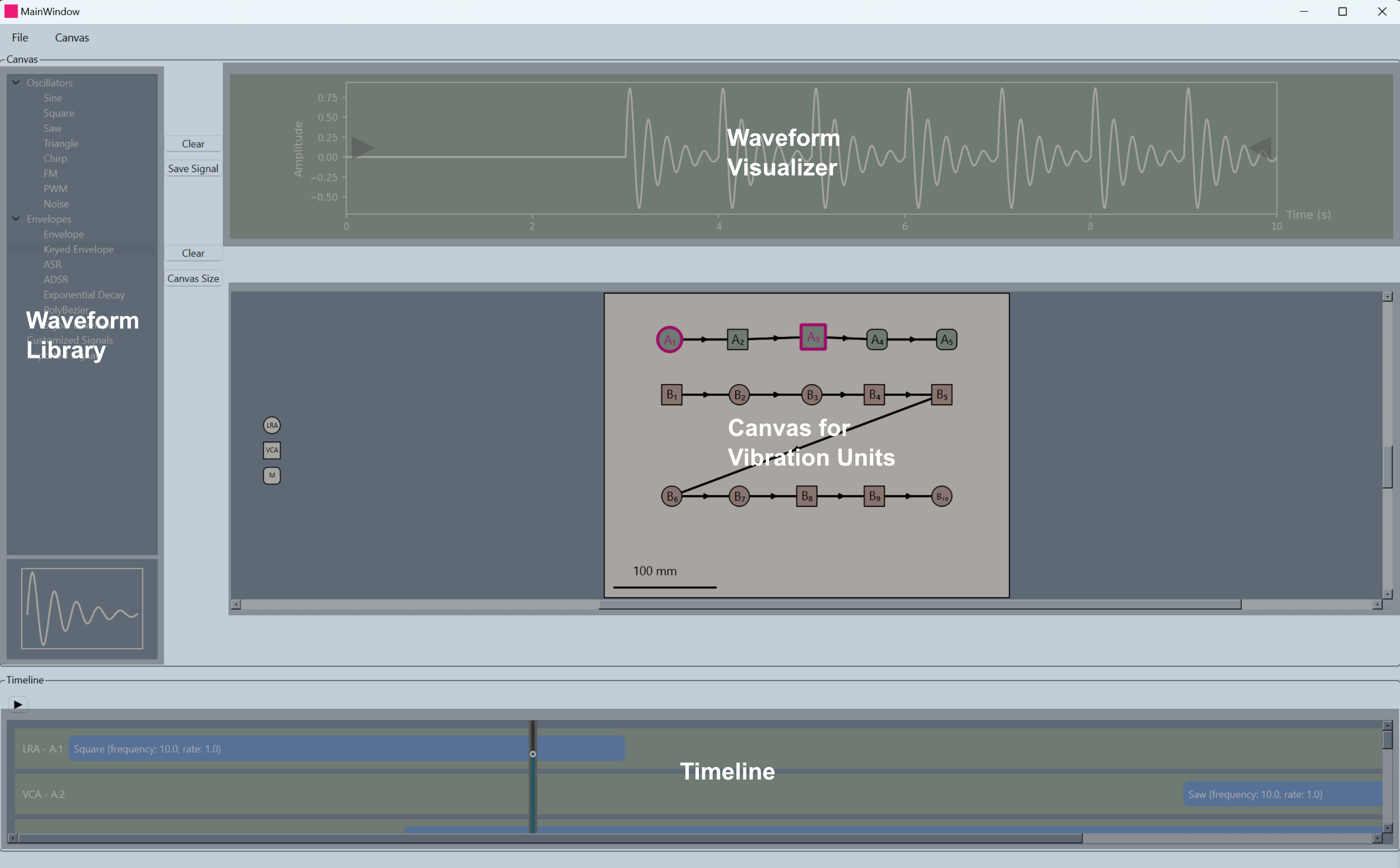}
    \caption{The layout of the GUI Editor, with its Waveform Library, Waveform Visualizer, Canvas, and Timeline.}
    \Description{The image shows a software interface with four main sections: The Waveform Library in the left panel contained a list of different waveforms available. The Waveform Visualizer at the top graphically depicted the selected waveform's shape and behavior over time. The Canvas for Vibration Units in the center are where vibration units were visually arranged and connected. The Timeline at the bottom displayed the sequence and timing of waveform actions and events.}
    \label{fig:sec3_gui_layout}
\end{figure}

\begin{itemize}
    \item \textbf{Waveform Library}: The Waveform Library included templates like oscillators and envelopes for creating custom waveforms. Designers could also import JSON waveforms from editors such as Syntacts \cite{pezent2020syntacts}, VIREO \cite{terenti2023vireo}, and Meta Haptics Studio \cite{oculus2023haptics}, enabling for the reuse of previous designs (DR3).
    
    \item \textbf{Waveform Visualizer}: The Waveform Visualizer allowed designers to create custom waveforms by combining templates from the Waveform Library via drag-and-drop. It also displayed the composed waveform for selected vibration units during pattern composition. 
    
    \item \textbf{Canvas}:
    The Canvas provides an approximate virtual layout of vibration units and their relative order within each chain. Designers could drag-and-drop a single unit into the Canvas. A "Create New Chain" function also allowed them to create a new chain of units at once and align the units in grid patterns. Selecting a unit would display its waveform in the Waveform Visualizer.

    \item \textbf{Timeline}: The Timeline summarized vibration patterns, with blue blocks indicating activation periods. Moving the slider highlighted active units on the Canvas. Clicking the "Play" button sent real-time vibration commands to the control unit via Bluetooth.
\end{itemize}

In a typical spatial pattern design workflow, designers would specify the number of chains and actuators for their system. The GUI then generated a default Canvas with the designated number of actuators, which could be freely adjusted to match their physical layout. Designers then assigned waveforms to each vibration unit, specifying start and end time. Waveforms can be created using templates from the Waveform Library or imported from other software. To review patterns, designers could click on a vibration unit to view its waveform in the Waveform Visualizer or use the Timeline slider to see a high-level summary of vibration patterns over time. This workflow was found to be easy to follow by both experts (Section~\ref{sec5_case_studies}) and novice users (Section~\ref{sec6:usability_study}).

To support the integration of VibraForge with external programming environments (DR3), we created a Python server and a Unity plugin that provided APIs to directly address the VibraForge hardware. The APIs offered functions such as \textit{SendCommand(address, start\_stop, intensity, frequency)} that enabled designers to configure the vibration parameters of a unit without delving into the waveform details, thereby assisting users without extensive technical expertise. The APIs were later used in Case Study 2 and 3 to build VR and robotics applications (Section \ref{sec5_case_studies}).

%% file: docs/4_technicalevaluation.tex
\section{Technical Evaluation} \label{section4:technical_evaluation}

% version 20240705

% \begin{itemize}
%     \item acceleration, shows precise intensity and frequency control (R4)
%     \item latency and bandwidth, to prove that the system is scalable. With X actuators, the latency is still under human perception threshold (R3)
%     \item acceleration - audio waveform, see if the measurement matches the expected waveform; see if the vibration can approximate audio-input (R4, R8)
%     \item simultaneous triggered actuators, voltage drop (R3)
%     \item repetitive vibration test, how long can a vibration unit last (R2)
%     \item temperature change? (R2)
%     \item power consumption, battery life (R6)
% \end{itemize}

% version 20240722

In this section, we demonstrate how the VibraForge toolkit fulfills the aforementioned design requirements via various performance metrics. For the evaluations, we used vibration units of X-axis LRAs and the small control unit with a 50\% intensity and frequency of 170 Hz, unless otherwise noted.

\subsection{Latency}

% another ref that could be useful: Temporal order and tactile patterns
Low latency and high bandwidth are crucial for data transmission in spatialized vibrotactile systems, particularly in scenarios where user actions trigger immediate tactile feedback, enabling users to respond and adjust their behaviors accordingly. To determine if VibraForge's latency was below human perceivable threshold of 45 ms \cite{vogels2004detection}, we measured the end-to-end latency of a PC to a control unit (i.e., latency of the Bluetooth Low Energy protocol (BLE)), and of a control unit to a vibration unit (i.e., latency of the UART). 

To measure the PC to control unit latency, a BLE GATT write command was sent and immediately followed by a read command. A Python timer computed the difference between the write command sending time and read command receiving time. We repeated this measurement 100 times and found the average round-trip transmission latency was $26.67 \pm 6.53 $ ms. Therefore, the one-way transmission latency was approximately $14$ ms. For the control unit to vibration unit latency, repeated measurements using an oscilloscope showed that each command took $125$ us to be transmitted from one unit to the next unit. Therefore, latency on a chain with 16 actuators would be $2$ ms. These added to an overall latency of approximately $16$ ms, which would be virtually imperceptible by humans \cite{vogels2004detection}.

\subsection{Bandwidth}

\begin{figure}[h]
    \centering
    \includegraphics[width=\linewidth]{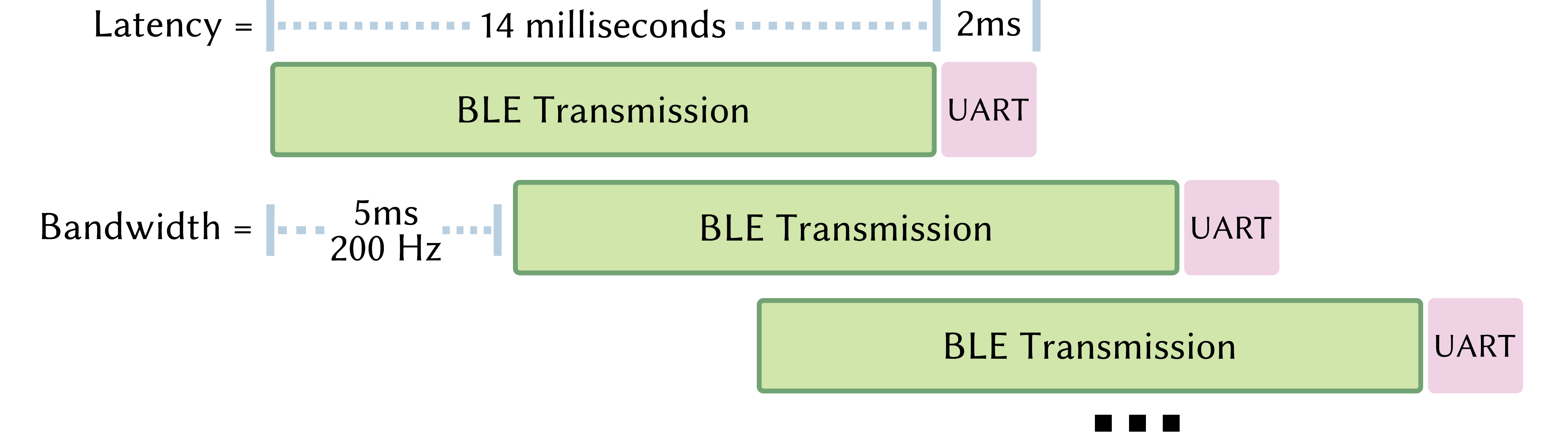}
    \caption{Data transmission bandwidth breakdown.}
    \Description{The breakdown of data transmission bandwidth and latency, highlighting BLE (Bluetooth Low Energy) transmission and UART communication. The BLE transmission segments were separated by 14 milliseconds, followed by 2 milliseconds of UART communication. The bandwidth was 5 milliseconds or 200 Hz, representing the frequency of the data transmission cycles. The diagram represents how BLE and UART transmissions alternate in time.}
    \label{fig:sec4_latency_bandwidth}
\end{figure}

VibraForge's bandwidth was defined as the number of commands that could be sent and processed per second by the control and vibration units. Bandwidth differs from latency in that asynchronous BLE commands could be continually sent without needing to wait for the previous command to be received (Figure \ref{fig:sec4_latency_bandwidth}). Since BLE commands (i.e., millisecond level) would be processed much slower than UART commands (i.e., microsecond level), the bandwidth bottleneck depended on the processing speed of the BLE commands on the control unit.

To measure VibraForge's bandwidth, a Python testing program repetitively sent BLE commands without latency and another program on the control unit received the commands and printed the receiving time to the serial port. The command processing time was the difference between the receiving time for each command. We sent $1,000$ BLE commands with five UART commands in each of them, and computed an average processing time of $2.96 \pm 1.20$ ms. Thus, during actual operation of the toolkit, the BLE command sending delay was set to $5$ ms with some tolerance. The control unit could simultaneously update five vibration units at $200$ Hz refresh rate, satisfying the signal segmentation algorithm requirements.

Although vibration units were not directly driven by continuous audio signals from the central MCU, the high bandwidth ensured that they could approximate continuous waveforms using discrete commands (DR2).

\subsection{Power Consumption and Battery Life}
Power consumption measurements provide an estimation of how long wearable devices would function before recharging was needed (DR5). Power consumption was measured using a bench power supply with a real-time measurement of voltage, current, and power. When only the control unit was connected, it consumed $106$ mA current. When multiple vibration units were connected to the control unit, each vibration unit consumed $2.5$ mA when idle. When a vibration unit was activated, the actuator consumed $150$ mA. Therefore, for a vibrotactile arm display with one control unit and two full chains of vibration units ($16x2)$, the idle current would be $186$ mA and a $500$ mAh LiPo battery would last $2.69$ hours. Even in extreme use cases such as multimedia entertainment, continually running two actuators would consume $486$ mA, leading to $1.03$ hours of operating time. For a larger system with four chains of vibration units ($16x4$), the idle current would be $266$ mA, with a $8200$ mAh LiPo battery lasting $30.83$ hours. In an extreme situation with eight actuators continuously activated, the overall current would be $1460$ mA and the battery would last $5.62$ hours.

\subsection{Maximum Number of Vibration Units On Each Chain}

Here we described how the maximum number of units on each chain was determined (DR1). Due to incremental wire resistance along the chains, voltage drops become more significant when additional vibration units are connected or activated simultaneously. To maintain stable operation, the MCU supply must maintain above $2.3$ V (i.e., the MCU's minimum operational voltage) and the actuator supply should not fall below $0.9$ V (i.e., LRA-X actuator's rated voltage). 

\begin{figure}[h!]
    \centering
    \includegraphics[width=\linewidth]{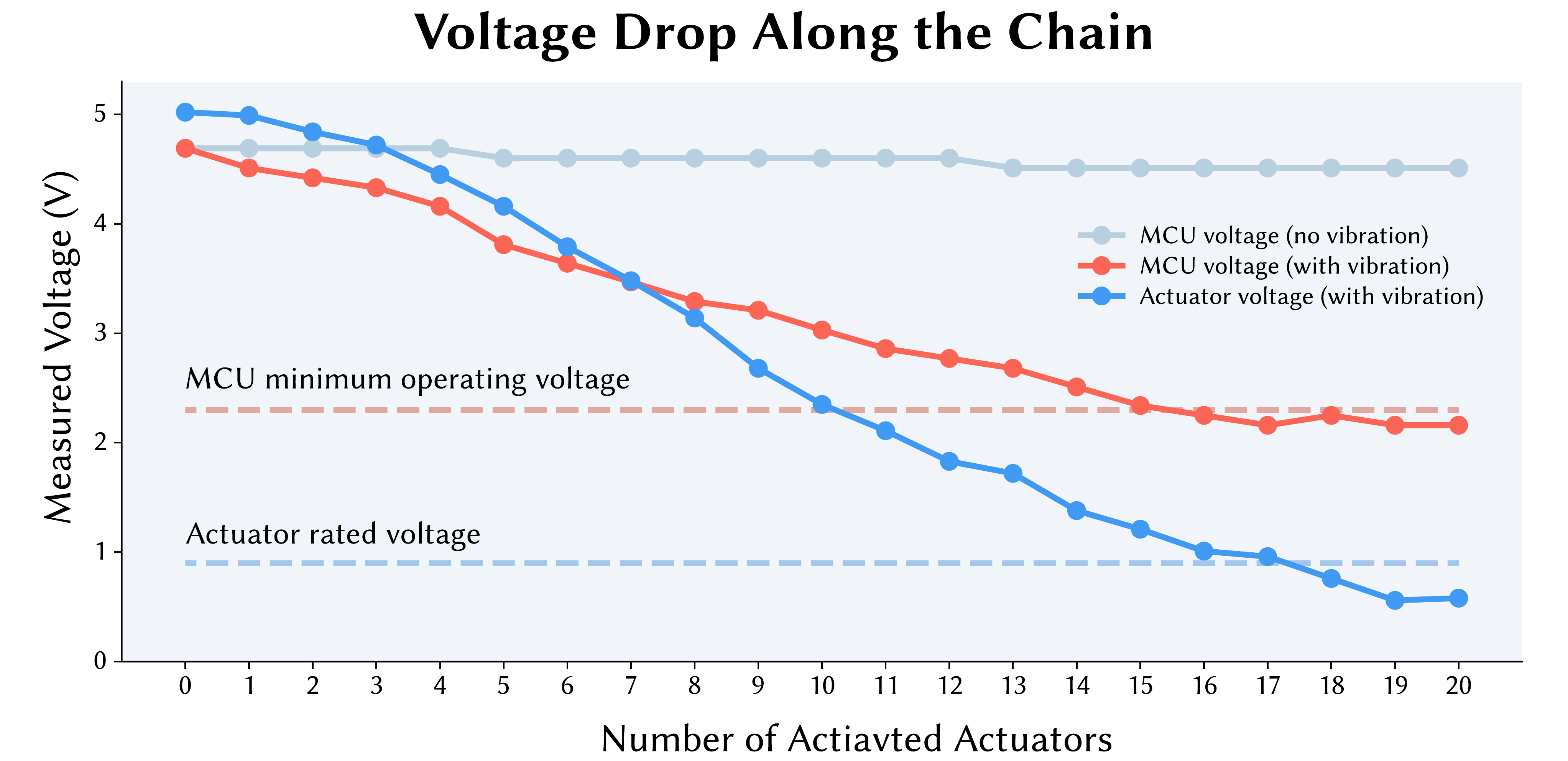}
    \caption{Results of the voltage drops on the MCU (red) and the actuator (blue) power supply with an increasing number of activated actuators.}
    \Description{This graph shows the voltage drop along the MCU and actuator power lines as the number of activated actuators increases, highlighting how the voltage decreases significantly for the actuators and slightly for the MCU, with critical thresholds marked for the MCU's minimum operating voltage and the actuator's rated voltage.}
    \label{fig:sec4_voltage_drop}
\end{figure}

We measured the voltage drop threshold by incrementally activating vibration units on a chain. An oscilloscope measured the MCU power line and actuator power line at the last vibration unit of the chain. If the last unit maintained a stable voltage, then one could assume the other units were stable as well. When the number of actuators surpassed 16, the MCU voltage dropped below the operating range (Figure \ref{fig:sec4_voltage_drop}). When it surpassed 17, the actuator voltage dropped below its rated voltage. Therefore, control units could support up to 16 actuators on each chain.

\subsection{Vibration Characteristics}

To demonstrate that the toolkit can provide precise control over different vibration characteristics (DR2), we measured the acceleration of vibration units at various intensities and frequencies.

\begin{figure*}[h]
  \centering
  \includegraphics[width=\linewidth]{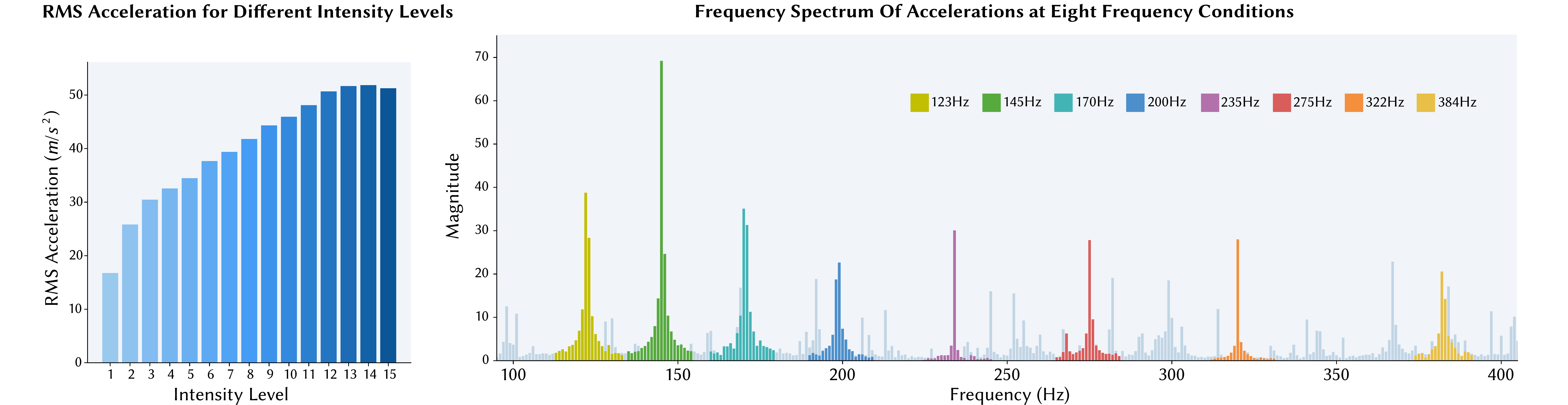}
  \caption{The acceleration measurements at various intensity and frequency levels.}
  \Description{This figure presents two measurements: on the left, a bar chart showing the increase in RMS acceleration as the intensity level rises from 1 to 15, and on the right, a frequency spectrum displaying the magnitude of accelerations at eight different frequency conditions ranging from 125 Hz to 384 Hz.}
  \label{fig:sec4_vib_chars}
\end{figure*}

For intensity, we used designed a FlexPCB version of the ADXL345 accelerometer and attach it between the skin and the vibration unit. A vibration unit was mounted on a sleeve and placed near the wrist of a human subject. During testing, the vibration unit was activated for three seconds at each intensity level and the vibration frequency was kept constant at 170 Hz. The average acceleration was measured as the RMS value over the 1-second period of stable vibration. Acceleration was found to increase monotonically as intensity increased (Figure \ref{fig:sec4_vib_chars} Left).

Similarly, for frequency, the vibration unit was placed near the wrist and activated for three seconds at each frequency level with an intensity of 25\% . The acceleration data was processed using a fast Fourier transform and the resulting main frequency responses were found well aligned with the designated frequency at each level, thereby demonstrating the precise PWM signal control (Figure \ref{fig:sec4_vib_chars} Right).

\subsection{Waveform Approximation}

\begin{figure*}[h]
    \centering
    \includegraphics[width=\linewidth]{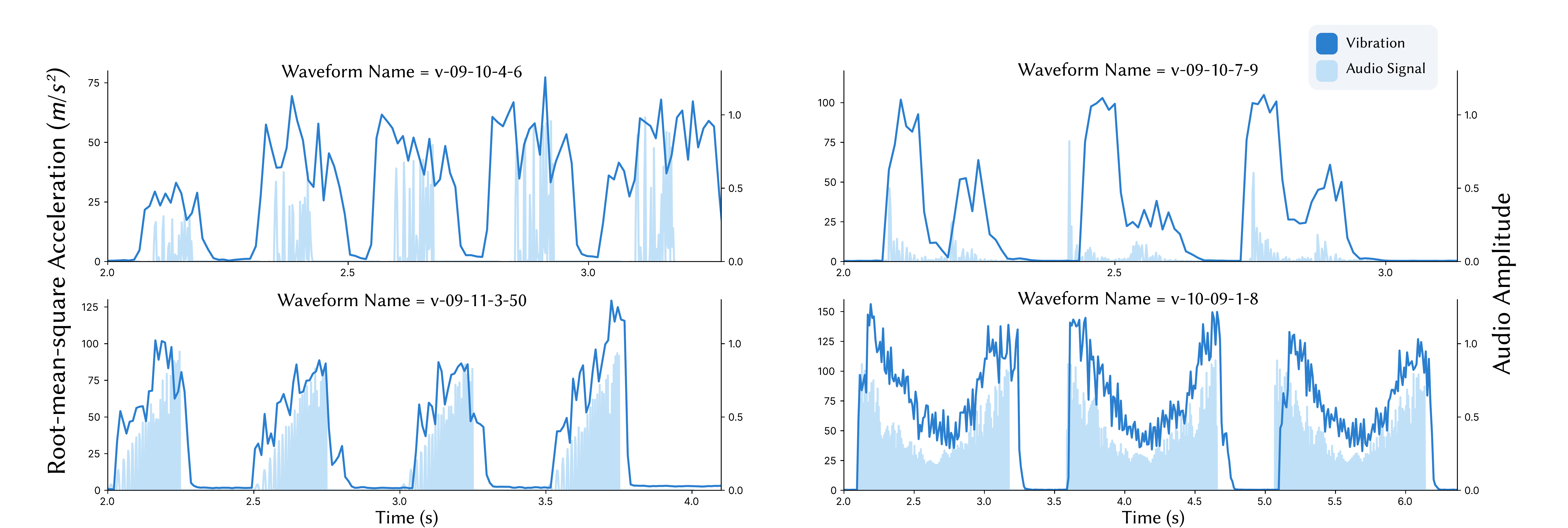}
    \caption{Comparison of actuator vibrations (blue lines) and original audio waveforms (light blue lines) from the VibViz Library.}
    \Description{This figure compares actuator vibrations (blue lines) and original audio waveforms (light blue shaded areas) from the VibViz Library, showing four different waveforms.}
    \label{fig:sec4_acc_command_pair}
\end{figure*}

Because vibration units are controlled via discrete UART commands rather than continuous audio signals, one might worry that this would reduce the expressivity of the vibration waveform delivered through the units. By utilizing high-bandwidth communication (DR2), VibraForge was capable of approximating audio signals and creating similar feedback. To demonstrate this, we used standard vibration waveforms from VibViz \cite{seifi2015vibviz}, converted them from WAV files to HAPTIC files using Meta Haptics Studio \cite{oculus2023haptics}, and sent them to the GUI Editor for conversion into a sequence of vibration commands to drive a vibration unit. After measuring the acceleration of the vibration unit and comparing it to the original audio envelope, both measurements closely approximated each other (Figure \ref{fig:sec4_acc_command_pair}).

\subsection{Cost Estimation}
VibraForge is a cost-effective toolkit. The small control unit costs \$27.23 USD (i.e., \$12.50 for the MCU, \$10.50 for the PCB, \$4.00 for the LiPo batteries, and \$0.23 for the 3D printed parts), which is similar to Arduino Uno but cheaper than Arduino Mega2560 and Raspberry Pi. The vibration unit costs \$3.46 USD (i.e., \$2.43 for the PCB, \$1.00 for the LRA, and \$0.03 for the 3D printed parts). For the wearable vibrotactile device in Case Study 1, for example, the total cost of a small control unit and 24 vibration units would be \$ 110.27 USD. This low manufacturing cost would enable designers to reproduce our toolkit with low financial burden. Note that these estimates are based on mass production (e.g., 10 control units and 100 vibration units), so prices may increase when manufacturing in smaller quantities. 

%% file: docs/5_casestudyphonemic.tex
\section{Case Studies}
\label{sec5_case_studies}

We conducted three case studies to investigate the potential “ceiling” \cite{ledo2018evaluation,myers2000past} of the toolkit's design space, measured by scalability and expressivity. The first case study demonstrated the toolkit's capability for high expressivity by replicating a phonemic display that requires complex vibration patterns. The second case study showcased high scalability by utilizing a larger number of units for VR interaction. The third case study found a balance between the two metrics, where parameterized vibrations were used to convey collision risks during drone teleoperation.

\subsection{Case Study 1: Replicating A Phonemic-Based Tactile Display}
\label{sec5:case_study_1}

% TODO: here I will describe how I use the toolkit to reproduce a vibrotactile sleeve device that uses 24 actuators to represent phonemes in speech communication for deaf people. I will show a comparison of the final prototypes. And I will highlight the benefits of using VibraForge, such as fast assembly, easy authoring of phonemic patterns, etc.
% More benefits: personalized layout with easy adjustable positions.
% Improvement of portability: "The display, which is wearable but tethered to equipment that has yet to be miniaturized,"

In Reed et al.'s work, a phonemic-based tactile display used 24 actuators on the forearm to deliver tactile codes of 39 English phonemes to deaf and hard-of-hearing users\cite{reed2018phonemic}. Phonemes were mapped to vibration patterns using frequency, waveform, amplitude, spatial location, and movement. Although the display was found to be useful, the building process was challenging. A desktop audio device (MOTU 24Ao) with three custom amplifier boards was needed to drive the 24 actuators independently, which made the system bulky and unusable in real-world situations. In this section, we demonstrated how the VibraForge toolkit could simplify the development process, enhancing usability and portability.

% XXX hardware dev workflow figure
\begin{figure*}[h]
    \centering
    \includegraphics[width=\linewidth]{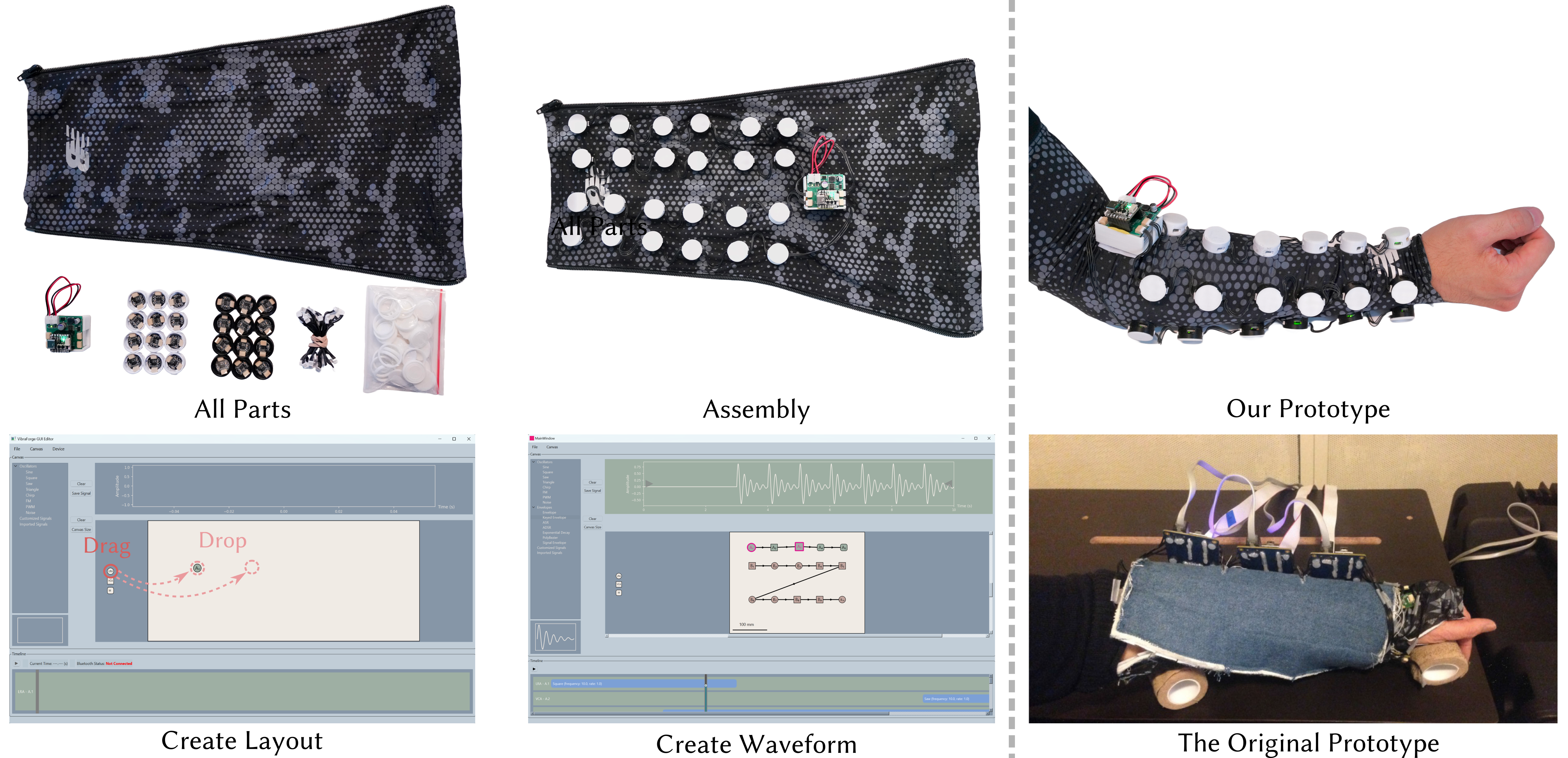}
    \caption{The development process of Case Study 1's phonemic tactile display re-implementation using VibraForge.}
    \Description{The top row shows the parts used for assembly, the assembly process itself, and the final prototype on an arm. The bottom row shows screenshots of waveform and layout creation, as well as an image of the original prototype for comparison.}
    \label{fig:sec5_phonemic_display}
\end{figure*}

\subsubsection{Vibrotactile System Development}
\label{sec:phonemic_display_development}

The first step was to determine the layout of the actuators and the chain connections. Since the original display used a 4x6 layout on the forearm, we decided to use four chains with six vibration units on each chain and a small control unit. Then we assembled the vibration units, attached them to the garment, and used JST wires to connect them (Figure \ref{fig:sec5_phonemic_display}). The control unit was then attached to the garment and four chains of vibration units were connected to the control unit. The mapping between chains and connection ports used the order of actuators described in the original paper. The assembly process took 16 minutes.

Once the hardware was built, we designed vibration patterns in the GUI Editor. Each phoneme was represented by a distinctive waveform that incorporated a high-frequency primary component and a low-frequency envelope, with a duration between 100 and 480 ms. Consonants were designed as static patterns with multiple actuators activated concurrently, whereas vowels were designed as dynamic patterns with actuators activated in succession. 

To design the waveform for the consonant "V" (IPA symbol /v/), for example, we began by dragging a 300 Hz sine wave into the  Waveform Visualizer and multiplying it with an 8 Hz sine wave, saving the result as a custom waveform in the library. If an envelope waveform like $cos^2$ (for consonant "H") was unavailable, we created it in Syntacts \cite{pezent2020syntacts} and imported it into the Editor. We then used the "create new chain" function to design a layout with four chains of vibration units on the canvas. We selected each of the designated units (i.e., i6, ii6, iii6, iv6), assigned the custom waveform, set the duration as 400 ms. Finally, we validated the design by clicking the "Play" button. An experienced haptic researcher averaged one minute and five seconds to design a phoneme, and would thus complete the entire process in 42 minutes. This streamlined design process would foster rapid design iterations.

\subsubsection{Lessons Learned}

The use of VibraForge to create the phonemic tactile display offered several advantages. First, it was fully portable (DR5) with a compact control unit and 24 vibration units integrated into the sleeve. Second, VibraForge streamlined the development process with solder-free module assembly and rapid waveform design in the GUI Editor (DR3). Third, its flexible design (DR4) enable easy position adjustments, addressing the design challenge of personalization mentioned in the original paper.

Meanwhile, some challenges remained. While the GUI Editor allowed for easy pattern creation, it mainly served as a testing site and lacked integration into real applications. If the designer wanted to play vibration patterns when certain phonemes are detected in speeches, the designer needed to manually convert patterns into sequences of vibration commands to fit the APIs. New API functions are needed to support the integration of GUI-designed waveforms into external programming environment.

%% file: docs/5_casestudyvr.tex
\subsection{Case Study 2: Fitness Gaming in Virtual Reality}

\begin{figure}[h]
\includegraphics[width=\linewidth]{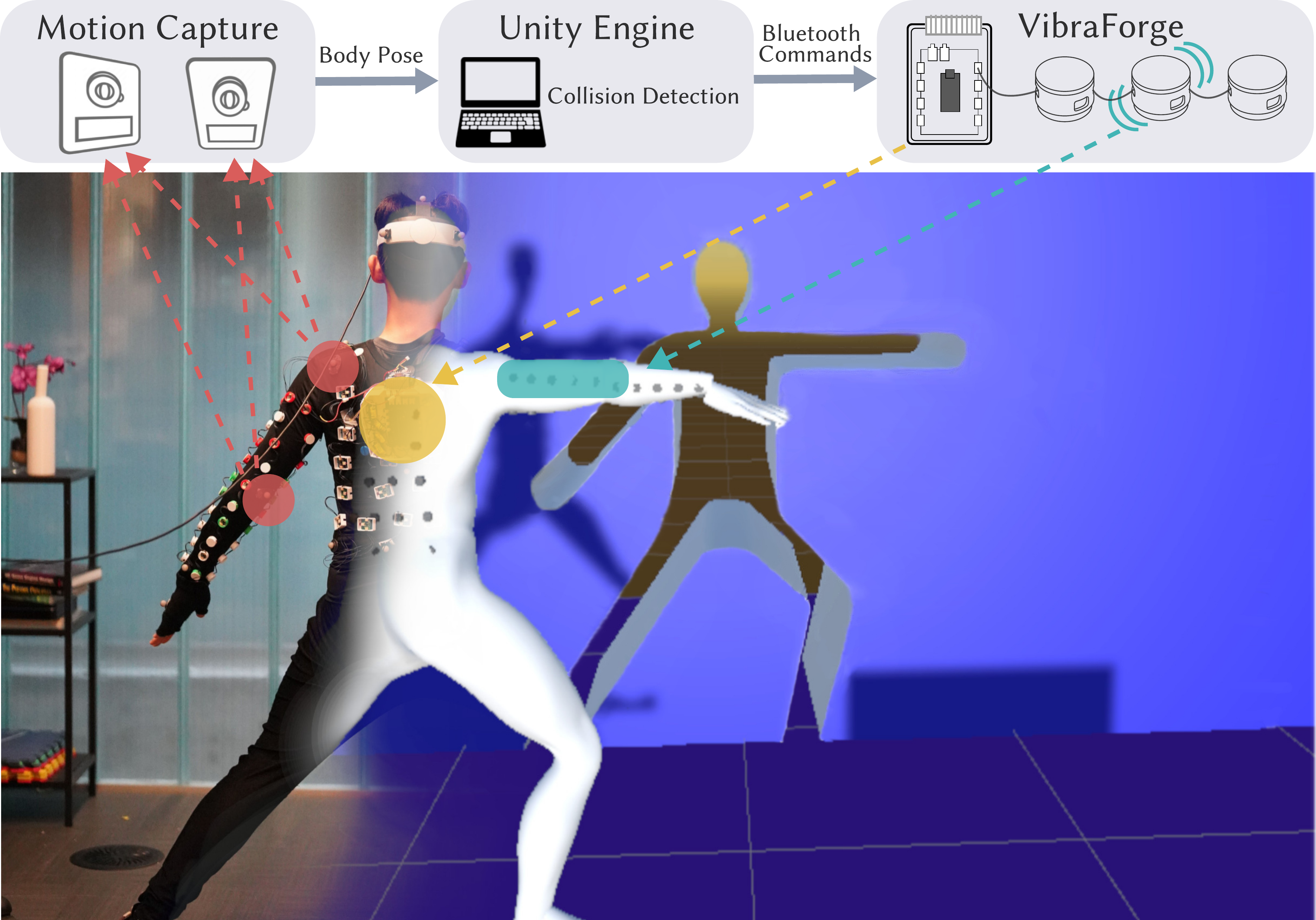}
\caption{An overview of the pipeline used in Case Study 2, where participant body movements were tracked by Vicon Motion Cameras and sent to a Unity scene for collision detection. Collision commands were then sent to a control unit on the participant's back through Bluetooth and converted to UART commands for the vibration units.}
\Description{The Vicon motion capture cameras captured one's motion and transmitted the body tracking data to a Unity scene, where it detected if a collision occurred between the participant and a virtual wall and activated the corresponding actuator to deliver tactile feedback.}
\label{fig:sec5_scene}
\end{figure}

Virtual Reality (VR) fitness applications like OhShape \cite{ohshape} and Supernatural \cite{supernaturalvr} have turned traditional fitness routines into interactive, gamified experiences. However, the absence of real-time and precise feedback about the physical positioning of one's body makes them difficult to use \cite{wu2023ar}. While prior work has demonstrated the utility of vibrotactile feedback for movement guidance \cite{bark2014effects,prabhu2020vibrosleeve,jo2023trainertap}, actuators were only deployed on major joint locations (e.g., wrists or elbows), thus offering limited and coarse guidance. Inspired by the TV show "Hole in the Wall"~\cite{holeinthewall}, we built a VR fitness application that asked participants to adjust their body to mimic cutouts in virtual walls (Figure \ref{fig:sec5_scene}). VibraForge was used to construct an upper-body suit that provided localized vibrotactile feedback to improve user performance. 

\subsubsection{Vibrotactile System Development}
The first step was to determine the suit layout and assemble the hardware components. We referenced prior work on spatial acuity \cite{elsayed2020vibromap,huang2024investigating} and determined that vibration units should be uniformly distributed on the arm with a spacing of 4 cm and on the body with an 8 cm spacing. The suit thus had 36 vibration units on the body and 40 on each arm (Figure \ref{fig:sec5_scene}). The large control unit was used to drive the substantial number of vibration units, and a compression shirt was used as the garment. We affixed the control unit to the upper back to help maintain balance. Then we affixed the vibration units and connected them with JST wires.

Once the suit was assembled, we integrated the hardware into the fitness game, which was built using the Unity Engine \cite{Unity2020_3_31}. The physical layout of the vibration units was manually converted into a virtual layout and overlaid on an avatar so that the  virtual units would follow human body movement. Whenever a collision was detected between a virtual unit and a wall, the system sent vibration commands to the control unit via Unity plugin APIs and activated corresponding units (Figure \ref{fig:sec5_scene}). Since collisions were binary (i.e., collided / not collided), the vibration intensity and frequency were set constant (i.e., 50\% and 170 Hz).

\subsubsection{User Study Design}

% Figure XXX OhShape player scene, including visual feedback / haptic feedback, and fusion of realistic and virtual scenes.

A preliminary study evaluated the effectiveness of the vibrotactile suit on fitness training performance. The VR application was run on an Oculus Quest 2 headset \cite{MetaQuest2} through a Quest Link connection to an Alienware M15 Laptop. Vicon Motion Capture cameras \cite{Vicon} tracked the body pose of each participant and mapped it to a virtual avatar in the Unity scene. Nine participants (i.e., 4 Males and 5 Females; mean age = 25 years, SD = 3 years) were recruited for the 1.5-hour study, and each received \$40 compensation. The study was approved by our institutional research ethics board and conducted in a university lab space. 

During the study, each participant would experience two conditions, i.e., no feedback and vibrotactile feedback. In the no feedback condition, participants only saw the incoming virtual walls and their own avatar and relied on proprioception to adjust their body position. In the vibrotactile feedback condition, participants would feel vibrations at the body positions where collisions occurred between vibration units and walls. For each condition, 24 pairs of walls with different cutouts were presented in sequence. Two walls in each pair were identical. The first wall served as the "preview wall" to help the participant adjust their body pose, and the second wall served as the "test wall" to measure any body position changes. We measured the number of collisions occurring during each condition. Our hypotheses were that (1) vibrotactile feedback would cause fewer collisions than no feedback, and (2) test walls would have fewer collisions than preview walls.

\subsubsection{Results}

% \begin{figure}[h]
% \includegraphics[width=0.60\linewidth]{figures/sec5_vr_collisions_cropped.pdf}
% \caption{The average number of collisions that occurred with and without vibrotactile feedback. The error bars depict the standard error.}
% \Description{Bar charts showing the results from the VR study.}
% \label{fig:sec5_results}
% \end{figure}

The average number of collisions between vibration units and walls was computed for each feedback condition. Then, a repeated measures ANOVA was conducted using SPSS to determine if feedback condition had an effect on the number of collisions that occurred. If there were significant differences between each feedback condition, pair-wise comparisons with a Bonferroni correction were then performed.

The RM-ANOVA determined that both feedback type ($F(1, 8) = 6.450, p < 0.05$) and wall type ($F(1, 8) = 5.582, p < 0.05$) had significant effects on the number of collisions. No interaction effect was found (($F(1, 8) = 3.936, p = 0.083$). For feedback type, the vibrotactile feedback (M=26.289, std=9.897) led to fewer collisions than the no feedback type (M=31.489, std=7.162, $p<0.05$). For wall type, the test walls (M=27.552, std=10) led to fewer collisions than the preview walls (M=30.226, std=7.694, $p<0.05$). Therefore, we concluded that both hypotheses were accepted.

\subsubsection{Lessons Learned}
Using VibraFogre to build the high-density vibrotactile suit offered several advantages. First, scalability (DR1) supported numerous vibration units (116 in total), providing localized feedback for guiding body movements. Second, the modular and flexible design (DR4) reduced assembly to less than 30 minutes with simple press-fit attachment and rapid adjustments. Third, low latency and high bandwidth communication (DR2) ensured real-time feedback when collisions happened, with no perceivable delays reported. 

Nonetheless, some challenges arose. The process of manually mapping the vibration units to the virtual layout in Unity was rather tedious. This could be solved by a localization pipeline that automatically detects vibration units' positions and their relative orders. Additionally, two participants reported inconsistent vibrations near the waist due to loosened fabric contact and body movement. We addressed the problem by temporarily tightening the suit, but future solutions are needed to ensure consistent skin contact. 

%% file: docs/6_casestudydrone.tex
\subsection{Case Study 3: Collision Avoidance for Unmanned Aerial Vehicle (UAV) Teleoperation}

\begin{figure}[h]
  \centering
  \includegraphics[width=\linewidth]{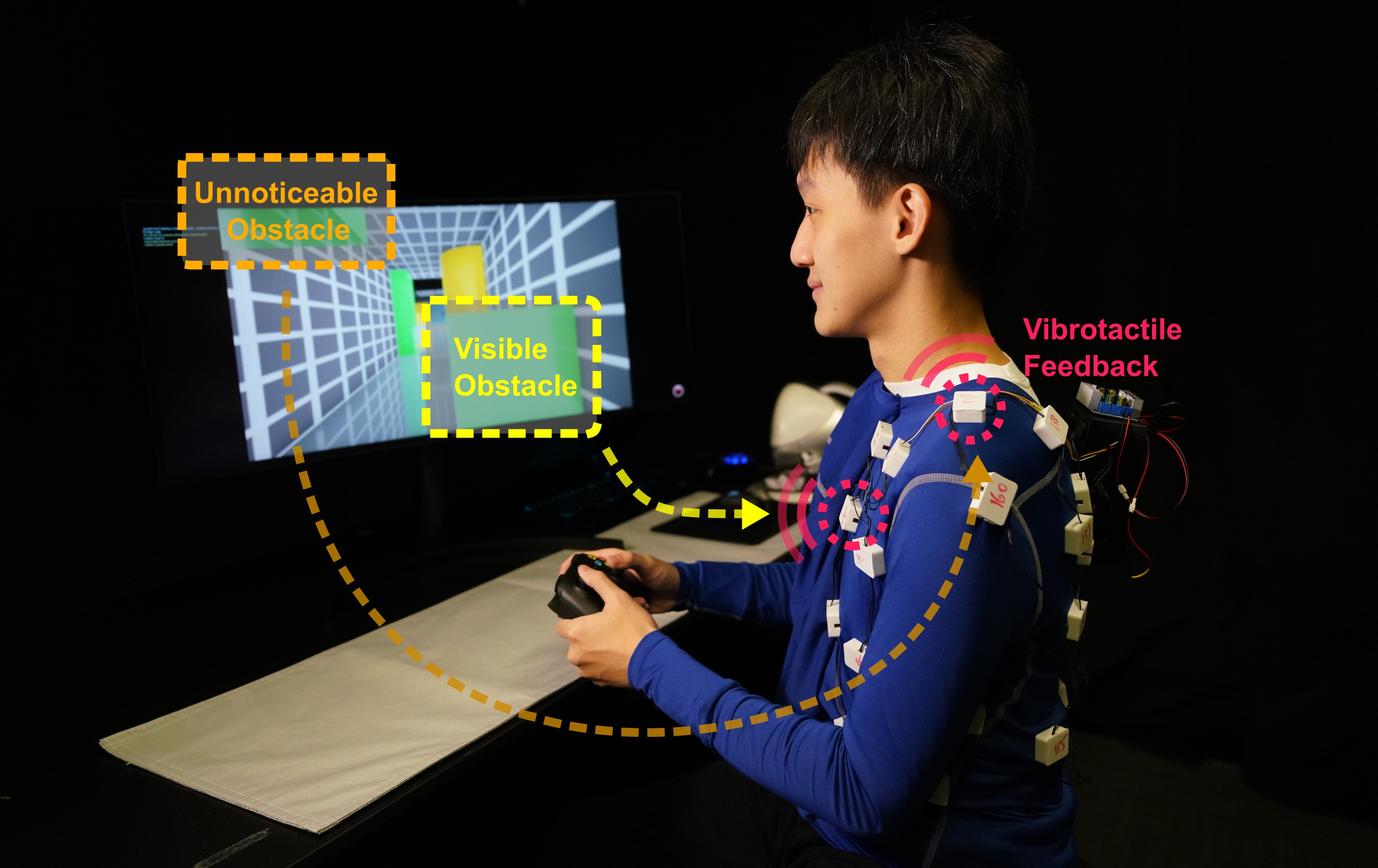}
  \caption{The vibrotactile device designed during Case Study 3 assists UAV operators with collision avoidance by delivering obstacle directions via multi-point bodily vibrotactile feedback. Operators not only see and feel visible obstacles (yellow), but also perceive obstacles outside their field of view (orange).}
  \Description{A human operator wearing the vibrotactile device is flying a simulated drone. He perceives incoming obstacles through vibrotactile feedback triggered on the body. He not only sees visible obstacles but also perceives obstacles outside the field of view.}
  \label{fig:sec6_uav_teaser}
\end{figure}

In UAV teleoperation tasks, operators heavily rely on visual feedback to detect obstacles and avoid collisions, but the narrow field of view from onboard cameras limits their situational awareness and increases collision risks. While prior work attempted to use haptic feedback for notifying operators of obstacles~\cite{brandt2009haptic,hou2017dynamic,zhang2020haptic}, they required operators to switch from radio control (RC) controllers to unfamiliar haptic joysticks, and provided only one directional force feedback. A spatialized vibrotactile feedback system using VibraForge could deliver simultaneous spatial cues about multiple obstacles, while operators continued using RC controllers (Figure~\ref{fig:sec6_uav_teaser}). Additionally, vibration characteristics can be used to convey different levels of risks, helping operators prioritize high-risk obstacles.

\subsubsection{Vibrotactile System Development}
\begin{figure}[h]
  \centering
  \includegraphics[width=0.8\linewidth]{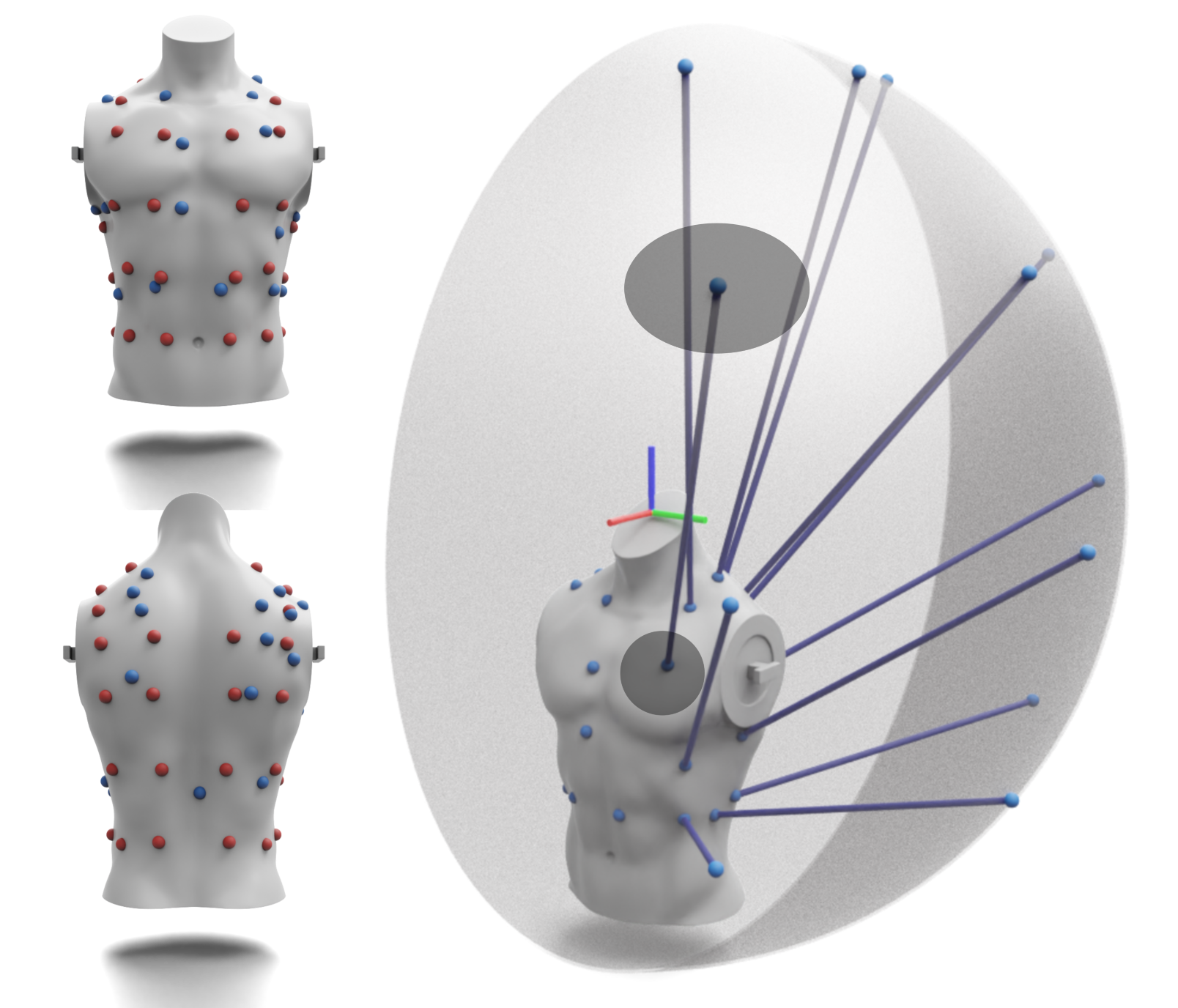}
  \caption{The layout of vibration units on the upper body. Red dots represent the initial layout with uniform spacing, while blue dots represent the final layout and are mapped to directions in the 3D space.}
  \Description{The layout of vibration units on the upper body. The initial layout was designed using 46 units with uniform spacing, while the final layout only used 32 units with more sparse spacing.}
  \label{fig:sec6_haptics_mapping}
\end{figure}

Our first step was to determine the layout of the vibration units on the body. Since there is no prior work on mappings between bodily tactile feedback positions and human perceptions of 3D directions, we conducted a perceptual study with 10 participants to identify a mapping. The layout was initialized with 46 vibration units uniformly distributed on the body and optimized using user data. The final layout used 32 vibration units to achieve comprehensive coverage of the 3D space (Figure~\ref{fig:sec6_haptics_mapping}). More details on the perceptual study and layout optimization can be found in \cite{huang2024aerohaptix}. Then we affixed press-fit 32 VCA vibration units to the garment, which was a compression shirt similar to Case Study 2. The vibration units were grouped into four chains, covering the front left, front right, back left, and back right. A large control unit was placed on the operator's back.

We then integrated the hardware into a Microsoft AirSim \cite{airsim2017fsr} simulation environment, built with Unreal Engine 4 and our Python server. Since the mapping between vibration units and 3D directions were known, when a potential collision was detected, the system would trigger the unit that had the nearest direction with the colliding object in the space. The collision risks with obstacles were computed using control barrier functions \cite{ames2019control} and scaled to $[0, 15]$ to match the actuators' intensity levels. If the risk of an obstacle was above $0.5$, the vibration unit in the corresponding direction would be activated by sending parameterized vibration commands through the Python server.

\begin{figure}[h]
  \centering
  \includegraphics[width=\linewidth]{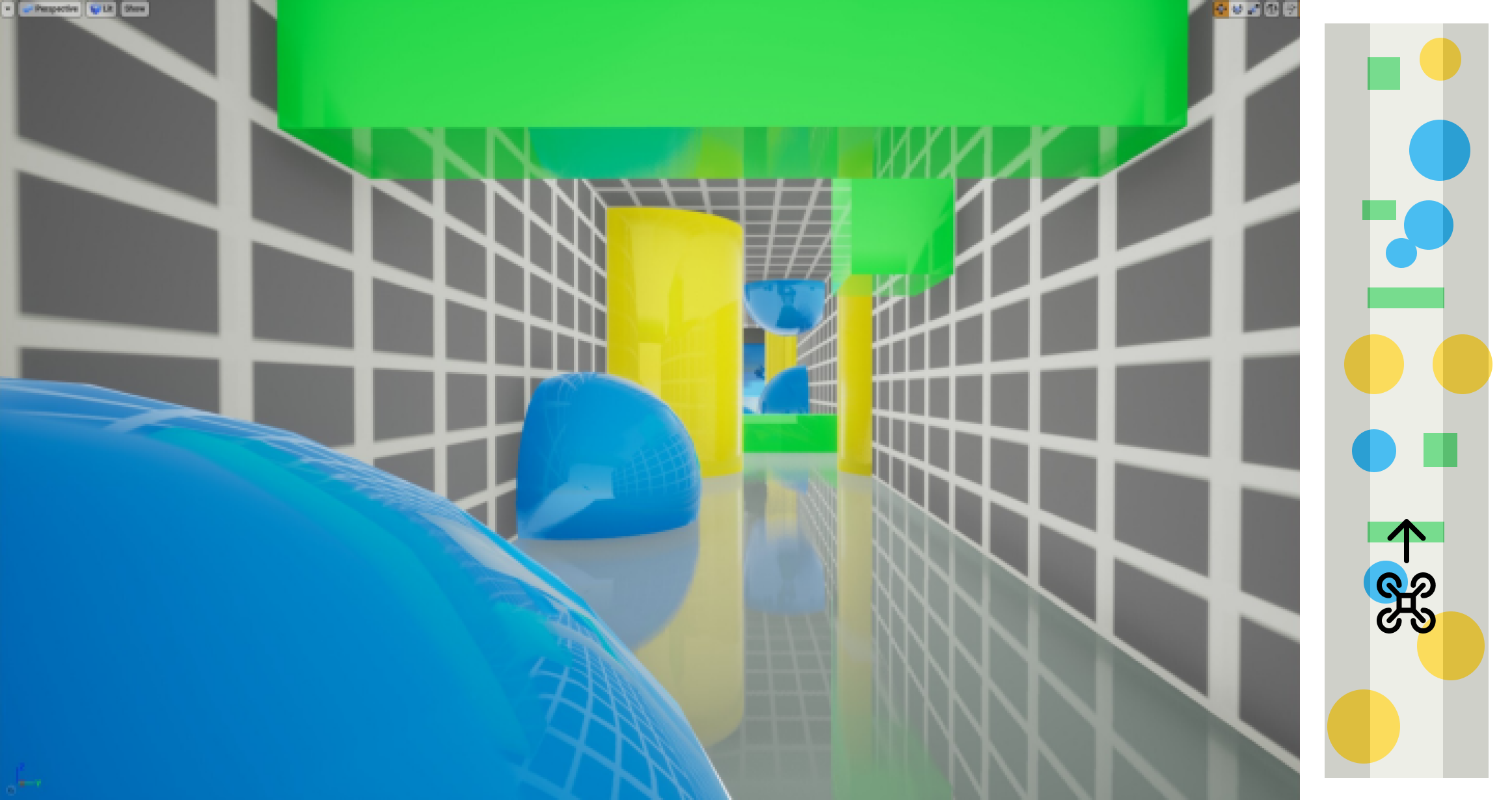}
  \caption{A first-person view of the simulation environment used in Case Study 3's study, with randomly positioned obstacles and a top-down view on the right.}
  \Description{This figure shows a navigation scene demonstrating the obstacle course where participants navigate using different control conditions.}
  \label{fig:sec6_map}
\end{figure}

\begin{figure*}[h]
  \centering
  \includegraphics[width=\linewidth]{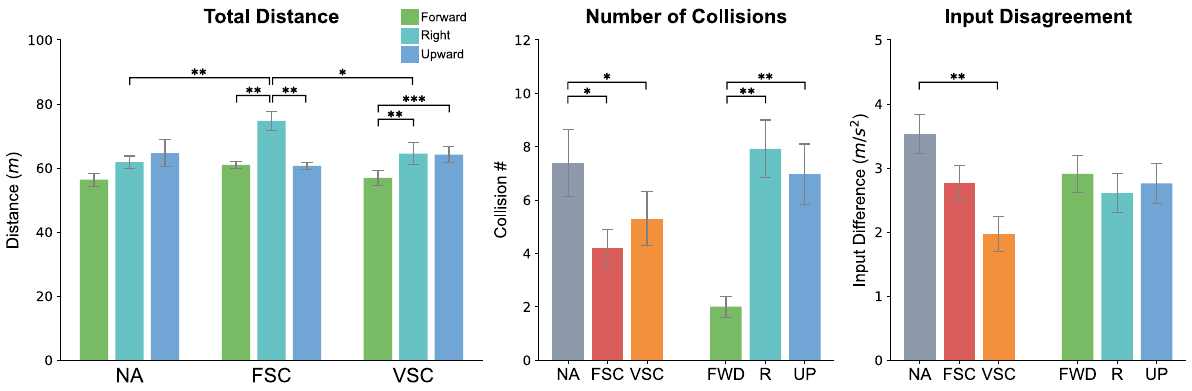}
  \caption{The results for the total distance traveled, number of collisions, and input disagreement metrics for the feedback conditions (NA, FSC, VSC) and flying directions (forward-FWD, right-R, and upward-UP). The error bars represent the standard error of the mean (SEM), $\oneS = p < 0.05$, $\twoS = p < 0.01$, and $\threeS = p < 0.001$.}
  \Description{Bar graphs display the performance metrics, including total distance, number of collisions, and input disagreement across multiple conditions (NA, FSC, VSC, etc.), highlighting significant differences between the control methods.}
  \label{fig:sec6_results}
\end{figure*}

\subsubsection{User Study Design}

A preliminary study was conducted to evaluate the effectiveness of the vibrotactile system on teleoperation performance. The simulation environment ran on an Alienware M15 laptop with RTX 3060Ti GPU. Twelve participants were recruited for the study (11 male, 1 female; mean age = 24 years, std = 3 years). Seven participants had previous experience operating quadrotors. Each study lasted 70 minutes and each participant received \$40 as compensation. The study was approved by our institutional research ethics board.

During the study, each participant went through three feedback conditions in a randomized order: no feedback (\textbf{NA}), force shared control (\textbf{FSC}), and vibrotactile shared control (\textbf{VSC}). For the NA and VSC conditions, participants used an Xbox controller for input, which had control mechanism similar to an RC controller. The output device for the VSC condition was the vibrotactile system. For the FSC condition, a Novint Falcon haptic joystick \cite{novint_falcon} was used for both input and output, similar to previous studies \cite{omari2013bilateral, reyes2015outdoor}. 

For each condition, the experimental scene was a $5 m\times5 m\times50 m$ tunnel (Figure \ref{fig:sec6_map}) that required participants to steer quadrotors forward, right, or upward. Different flying directions were used to vary visual information capacity as obstacles are often outside the camera's field of view when flying right and upward. Fifteen objects were randomly placed in the tunnel to avoid learning effects. Based on these conditions, we hypothesize that (a) vibrotactile feedback would enhance teleoperation performance compared to no feedback, and (b) the improvement would be more effective in the right and upward directions than the forward direction.

\subsubsection{Analysis and Results}

% metrics
We evaluated three metrics: total distance traveled, number of collisions, and input disagreement (i.e., the deviation of user steering commands from the safe input range during flight). Using a two-way repeated measures ANOVA in SPSS, we assessed the effect of feedback condition and flying direction. Post-hoc pairwise comparisons with a Bonferroni correction were conducted for significant results (Figure \ref{fig:sec6_results}).

The RM-ANOVA determined that feedback condition had a significant effect on total distance traveled ($F(2,22) = 3.585, p < 0.05$), as did with flying direction ($F(2,22) = 17.868, p < 0.001$). The interaction between feedback condition and flying direction was also significant ($F(1.957,21.527) = 4.916, p < 0.01$). The post-hoc analysis of the interaction effect revealed that FSC-R had longer distance than FSC-FWD ($p < 0.01$), FSC-UP ($p < 0.01$), NA-R ($p < 0.01$), and VSC-R ($p < 0.05$), while VSC-FWD had a significantly shorter distance than VSC-R ($p < 0.01$) and VSC-UP ($p < 0.001$).

For the number of collisions, the RM-ANOVA determined that feedback condition had a significant effect ($F(2,22) = 8.095, p < 0.01$), as did flying direction ($F(2,22) = 15.653, p < 0.001$). The interaction between feedback condition and flying direction was also not found to be significant ($F(2.168,23.846) = 1.910, p = 0.168$). The pairwise comparisons between feedback conditions revealed that NA caused more collisions than FSC ($p < 0.05$) and VSC ($p < 0.05$). The pairwise comparisons of flying directions revealed that flying forward caused fewer collisions than right ($p < 0.01$) and upward ($p < 0.01$).

For input disagreement, the RM-ANOVA determined that feedback condition had a significant effect ($F(2,22) = 4.798, p < 0.05$), while flying direction did not ($F(1.254,13.789) = 0.628, p = 0.476$). The interaction between feedback condition and flying direction was also not significant ($F(4,44) = 1.566, p = 0.200$). The pairwise comparison between feedback conditions revealed that there was less disagreement with VSC compared to NA ($p < 0.01$)

In summary, VSC caused fewer collisions and less input disagreement than NA, and achieved comparable performance with FSC. Additionally, the right and upward directions had more collisions than the forward condition, highlighting users' increasing reliance on haptic feedback when visual capacity was limited. Thus, our hypotheses were accepted.

\begin{figure*}[h!]
    \centering
    \includegraphics[width=\linewidth]{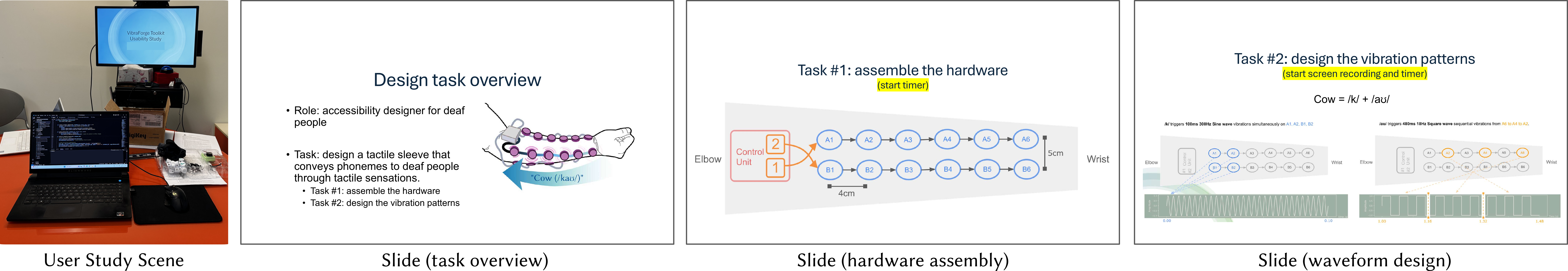}
    \caption{The usability study environment and examples of the diagrams provided to participants during the study.}
    \Description{The usability study environment, with a laptop, a mouse, a box of hardware components, and a display. Three slides used in the study showing diagrams of the design task. A task overview with a sketch illustration of the tactile sleeve, a sleeve layout guiding the hardware assembly, and two waveform illustrations for the vibration patterns.}
    \label{fig:study_scene}
\end{figure*}

\subsubsection{Lessons Learned}
Using VibraForge to build the teleoperation vibrotactile system revealed several benefits. The flexible design (DR4) enabled rapid iterations of vibration unit layouts, from an initial 46-actuator setup to a final 32-actuator configuration, allowing for quick testing of various spatial mappings. Low-latency transmission (DR2) provided immediate feedback essential for high-speed tasks, while varied intensities (DR2) offered nuanced information, helping participants prioritize high-risk obstacles. These features highlight the toolkit's capability for complex applications.

However, design challenges did arise. Participants noted that intensity perception varied across body areas, with stronger sensations near bones and weaker ones in areas with more tissue. Thus, individual calibrations might be needed to ensure feedback consistency. Additionally, the sparse final layout required soldering longer JST wires to replace the original ones. This suggested that VibraForge needs more enhancements to support sparse layout designs.

%% file: docs/RR_usabilitystudy.tex
\section{Usability Study}
\label{sec6:usability_study}

To understand the usability of the toolkit and determine its "threshold" (i.e., a user’s ability to get started using the toolkit \cite{ledo2018evaluation,myers2000past}), we conducted a first-use lab study \cite{hartmann2006reflective} with ten participants recruited from a local technology exhibition. Participants included five males and five females with ages ranging from 18 to 40 years old (Mean = 25.6 years, SD = 6.7). Participants had a range of prior experience programming, building physical devices, and building haptic/tactile devices, with some never having done so and others proclaiming themselves as experts. Each study lasted one hour and each participant received \$30 as compensation. The study was approved by our institutional research ethics board.

\subsection{Study Design}

The study consisted of a design task, a free exploration activity, and a post-study questionnaire and interview. In the design task, participants replicated a simplified version of the phonemic tactile sleeve display from Case Study 1 \cite{reed2018phonemic}: assembling two chains with six vibration units each and designing two phonemes, consonant /k/ and vowel /au/).  Participants received a 5-minute tutorial on the toolkit's hardware components (i.e., control units, vibration units, connector direction, garment attachment, etc.) before assembling the tactile device following a diagram displayed on a computer monitor. After assembly, participants underwent another 5-minute tutorial on the GUI Editor and waveform creation workflow. They then designed and tested the two phonemes using diagrams on the computer monitor. 

For the 15-minute free exploration activity, participants could assemble new devices on other garments (e.g., headband, glove, socks, etc), or combine multiple waveforms to produce complex patterns in the GUI Editor. Participants were encouraged to utilize their creativity, so the experimenter did not provide specific guidance but rather served as an assistant helping participants achieve their design goal. The experimenter recorded the creation process and the final outcome from the exploration. After that, participants completed a questionnaire with demographic questions and questions from the System Usability Scale (SUS) \cite{brooke2013sus}. A semi-structured interview then focused on the overall design experience, key areas for improvement, and potential applications of VibraForge.

\subsection{Results}

All participants successfully completed the design task and free exploration activity. The average completion time for the design task's hardware assembly was 8 minutes 39 seconds (SD = 1 minute 52 seconds) and 8 minutes 3 seconds (SD = 2 minutes 46 seconds) for the design task's software waveform design. These completion times were comparable to the expert time reported in Section \ref{sec:phonemic_display_development}. Furthermore, an average SUS score of 76.75 was obtained, which is between the 70th and 80th percentile and higher than average SUS scores \cite{brooke2013sus}. These results indicate that the toolkit has a very low "threshold" for participants.

% strengths of the toolkit
Overall, participants appreciated the intuitive design of the hardware components and found the assembly process straightforward and intuitive (e.g., P10: \textit{"the hardware was very easy to assemble and it was very intuitive to configure"}; P2: \textit{"it feels like LEGO pieces clicking together"}). Participants also appreciated that the software was easy to learn and demonstrated good interface design (e.g., P1: \textit{"the software part was nice because it was similar to the designing tools like Canva or Figma"}; P6:  \textit{"the timeline at the bottom was very helpful for me when I was trying to debug what was going on"}).

\begin{figure}[h!]
    \centering
    \includegraphics[width=\linewidth]{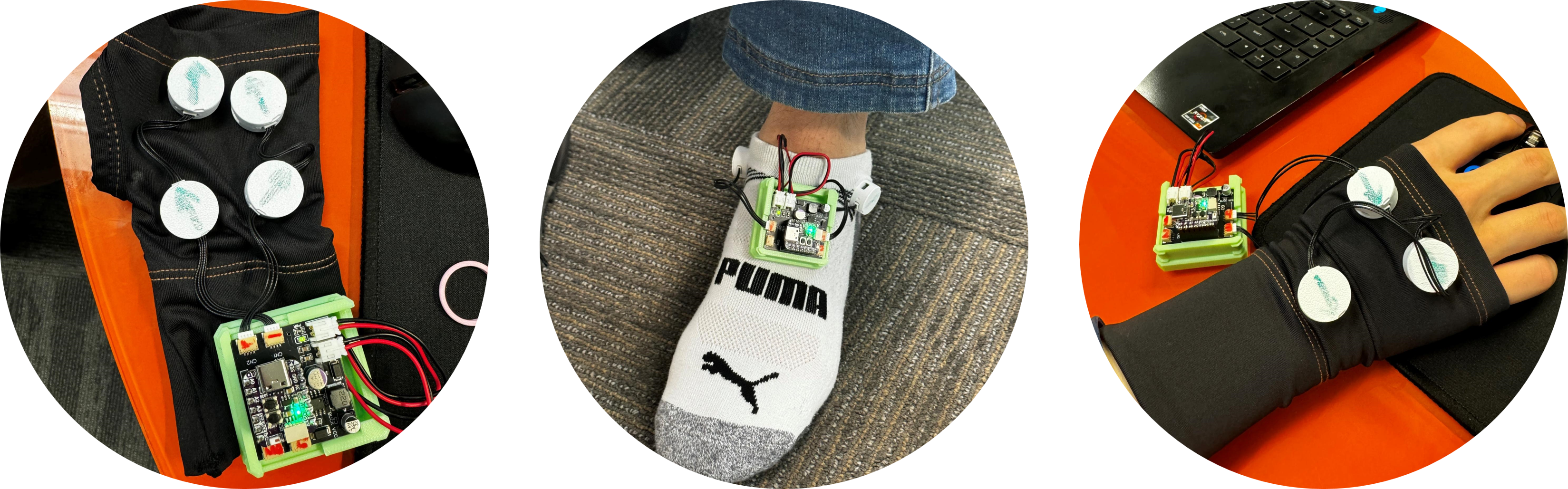}
    \caption{The devices created by participants during the free exploration: a glove with 4 vibration units on the dorsal side (P4), a sock with two units on the left and right side of the heel (P7), and a glove with three units on the dorsal side and one unit on the palm for transcribing songs (P10).}
    \Description{The devices created by participants during the free exploration activity: a glove with 4 vibration units on the dorsal side (P4), a sock with two units on the left and right side of the heel (P7), and a glove with three units on the dorsal side and one unit on the palm for transcribing songs (P10).}
    \label{fig:free_exploration_devices}
\end{figure}

During the free exploration activity, participants created a variety of haptic devices (Figure \ref{fig:free_exploration_devices}). For example, P4 designed a glove with four vibration units on the dorsal side of the hand. P7 attached two vibration units on the left and right sides of the heel on a sock to notify no-entry area. P10 designed a glove with three units on the dorsal side and one unit on the palm to "play" the Queen song "We Will Rock You" where each note triggered a different unit on the hand. P2 used the GUI Editor to transcribe the Nat King Cole song “On a Bicycle Built for Two” and played it using the frequency and duration of notes on each vibration unit. P8 made attempts to reproducing other phonemes such as /sh/, /ei/ and /p/ from the original paper \cite{reed2018phonemic} using the GUI editor. 

% free exploration, show the variety

Participants did note some areas for improvement. For the hardware components, participants mentioned that it would be helpful to have more apparent visual feedback on the chain connections (e.g., P2: \textit{“I feel like it [chain connector index] would be much more visible if it's printed on top of it rather than in the middle.}; P7: \textit{“I didn't know whether they [vibration units] were connected ... I think making [the LED] more visible would be more helpful"}. They also suggested other methods to affix the vibration units to garments to make the toolkit more adaptable to other fabrics (e.g., P7: \textit{"This [bottom ring attachment] and a magnet would have been a bit more effective, because you would only need to press it ... and it's gonna have a really strong grip on it for you to move it around"}; P4: \textit{"maybe there's like a lock method so, when you snap on it locks into the sensors [vibration units]"}. Based on these feedback, there is thus a need to add visual indicators for hardware connection statuses and explore additional mechanical affixation structures.

Others expressed the need for more versatile functionality in the GUI Editor. For example, P1 wanted to directly manipulate the waveforms (e.g., \textit{"if I want to change the amplitude, for example, I should just drag the top part and then drag down to lower, like direct manipulation ... if I want to change the frequency, I should just click and drag the amount to left or right"}. Similarly, because participants needed to drag and drop the same waveform onto several vibration units, P6 suggested adding "multi-select" functions to edit multiple vibration units concurrently. Because the Waveform Visualizer was used to both edit and assign waveforms, some participants were confused about its current status. P5 commented that \textit{"the fact that the designing of a custom wave and assigning it to the node share the same interface is a bit confusing"}. These comments suggest that the GUI Editor requires some refinement to ensure that supported actions are more natural and clearly conveyed to users.

Participants also highlighted several potential applications of the toolkit. P1 and P4 noted how VibraForge could be used to aid visually impaired individuals with navigation. P2 and P10 demonstrated how the toolkit could be used to translate music into tactile patterns, whereas P6 suggested using it to enhance VR and gaming immersion by simulating action feedback. P3 and P9 proposed the creation of devices for personalized massages and stress relief. P7 was enthusiastic about using it for safety applications, such as non-visual notifications while running or cycling, particularly in noisy environments. These examples demonstrate VibraForge’s potential usage in diverse domains.

%% file: docs/7_discussion.tex
\section{Discussion}

Herein we discuss the strengths and limitations of VibraForge summarized from the studies, and suggested potential solutions to address the limitations. Additionally, based on our first-hand developing experiences, we synthesized a generalized five-step pipeline for building spatialized vibrotactile systems.

% what case studies tell us about VibraForge?
\subsection{Strengths of VibraForge}

VibraForge aims to provide a "least resistance path" \cite{ledo2018evaluation} towards creating spatialized vibrotactile systems via bridging the gap between scalability and expressivity. The systems designed in case studies ranged from a sleeve-type phonemic display on the forearm, to a body-scale system spanning the upper body. Compared to existing toolkits that are often constrained to fewer than 16 actuators~\cite{wittchen2022tactjam,pezent2020syntacts,dementyev2021vhp}, VibraForge's chain connection method enables designers to easily scale their system to more than 100 vibration units, significantly expanding potential design spaces. Furthermore, new units can be plug-and-play into existing systems without additional programming, fostering rapid prototyping process. VibraForge also features great expressivity. It rendered tactile information at different complexities, ranging from simple on/off status for VR collision detection, to parameterized risk levels in teleoperation and symbolic phoneme representation in the phonemic display. Compared with previous spatialized vibrotactile systems that were restricted to using ERMs and PWM control~\cite{Wang2021Research,OpenVNAVI,lemmens2009body}, our toolkit offers various actuator options and provides fine-grained control over vibration characteristics, thus reliably reproducing complex waveforms.

% talk about the GUI editor using usability study.
Furthermore, VibraForge demonstrates high usability. Case Study 1 and the usability study indicated that both professional and novice designers benefited from the streamlined hardware assembly and intuitive GUI design. Compared with prior design interfaces~\cite{pezent2020syntacts,minamizawa2012techtile}, our GUI Editor provided substantial support for multi-actuator pattern design. Professionals using the GUI Editor effectively reduced the average pattern authoring time. Novice users compared the intuitive hardware assembly to the joyful process of "building LEGOs". Within an hour, they learned the toolkit basics and utilized it to finish professional design tasks and for creative exploration. This indicated that the toolkit lowered the technical barrier to building spatialized vibrotactile systems and enhanced the overall efficiency.

VibraForge also features portability and flexibility. Compared with the original phonemic display that was wired to a bulky desktop audio device, our components were more portable and easy to wear, such as the small control unit that communicated via Bluetooth. In the UAV teleoperation, flexibility ensured that vibration units could be frequently adjusted during the layout optimization perceptual study. Compared to using commercial devices~\cite{bHapticsWebsite} that often have fixed feedback positions, VibraForge enables designers to easily adjust the positions of vibrations units, thus enabling rapid design iterations.

In summary, rather than contributing one incremental technical advancement, our work fuses multiple components—modularity, usability, and portability—into a cohesive vibrotactile prototyping toolkit. This holistic system-level integration enables a larger design space with a higher “ceiling” (shown in the case studies) and lower “threshold” (shown in the usability study), opening new possibilities for both professional designers and novice users to build applications with haptic feedback.

VibraForge could be useful within several domains. For example, In VR social platforms, spatial vibrotactile feedback could help users express their emotions through social touch interactions \cite{sykownik2020experience}. Developers using the toolkit can easily adjust the layout of actuators to build personalized devices, in order to avoid discomfort or sensory issues in certain body areas. As another example, VibraForge can be used to build prototypes for tactile perceptual studies, such as those measuring spatial acuity, information transfer, or other perceptual characteristics. For example, Park et al. performed perceptual studies to measure information transfer across five body positions \cite{park2023information}, but the choice of positions were limited by the layout of bHaptics modules~\cite{bHapticsWebsite}. By using VibraForge, researchers could freely adjust the layout of vibration units on the body and adapt it to any study requirements.

\subsection{Limitations of VibraForge}

The toolkit does not come without limitations. The open-loop control used in the chain connection poses some issues. For example, when a control unit sends a vibration command, it can be unclear if the command was successfully transmitted to the target vibration unit. When transmitting a "stop" command, this is especially important as a vibration unit will not stop vibrating if the commands was missed. Additionally, when multiple actuators are activated there is a noticeable voltage drop along the chains, which would affect users' experience of the vibrations. By adding extra command receiver pins to the control unit, the vibration and  control unit would form a loop, wherein the last vibration unit in a chain would be connected to the control unit using a receiver pin and could propagate a confirmation message back to the central unit. In cases where commands fail during transmission, the central unit could resend the message. A closed loop connection would also reduce the voltage drop along the chain, as both ends of the chain would be connected to the power source. An LTspice simulation showed that an open-loop, eight-unit chain would result in voltage drop from 5 V to 3.1 V, while a closed-loop would increase the voltage to 4.2 V.  

While the GUI Editor's waveform design was advantageous for the phonemic display's pattern design, if a designer wanted to create an effect where actuators vibrated out from a central point in concentric circles, a designer would have to manually assign the waveform and specify the start and end time for each vibration unit. Instead, graphical design metaphors could be used to break down spatial patterns into low-level animations that would be automatically converted to vibration commands for each unit. 

VibraForge was built focusing on dense matrices of vibration units, as this was consistent with prior spatial tactile feedback systems that employed dense actuator arrays (e.g., facial masks \cite{Wang2021Research} and bodysuits \cite{bHapticsWebsite,lemmens2009body, west2019design}). However, Case Study 3 revealed that it was challenging to support the sparse layout of vibration units on the body, as longer wires were needed to connect more distant units, which may impact the overall system reliability. To overcome this limitation,  we plan to support the use of multiple control units within the same system so that designers can place control units at various positions on the body to support such sparse layouts.

Our studies also emphasized the need for an automated pipeline to visualize vibration unit positions. Currently, the GUI Editor canvas provides an approximate virtual layout of vibration units and their relative order within each chain. Although participants in our usability study were able to map physical units to their virtual counterparts, a true-to-scale, integrated visualization would reduce cognitive load and foster more rapid design iterations. One potential solution could be to revise PCB designs and add programmable LEDs to vibration units. When units are activated sequentially on each chain, computer vision algorithms could detect the positions of units on the body and their relative orders. Alternatively, ultra-wideband (UWB) tags could be mounted on top of the vibration units. With multiple UWB anchors setup in the environment, each UWB tag could utilize trilateration to find its own position. We are actively exploring these solutions and will incorporate them into the next iteration.

Lastly, VibraForge currently only supports vibrotactile actuators. Recent multimodal haptic toolkits, such as SleeveIO \cite{shtarbanov2023sleeveio} and TactorBots \cite{zhou2023tactorbots}, demonstrated the potential of integrating diverse actuator types (e.g., pneumatic, hydraulic, servo-driven) into haptic systems. However, these toolkits often rely on direct connections between modules and controllers, which limit their scalability. In contrast, VibraForge’s chain-connection architecture naturally supports extensions. Adding new actuator modalities would simply require custom PCB drivers and revised data protocols. Data bytes for intensity and frequency could be repurposed to handle parameters such as servo speed and position, Peltier temperatures, mechanical displacements, and so on. Thus, VibraForge could achieve the expressivity possible with other multimodal toolkits while maintaining its scalability.

\subsection{Pipeline for Creating Spatialized Vibrotactile Systems}

Synthesizing the processes followed during the case studies, we summarize a generalized five-step pipeline for creating spatialized vibrotactile systems. The first step is to determine \textbf{what} information to deliver using vibrotactile feedback. For example, the purpose of the phonemic display was to convey vowels and consonants to deaf and hard-of-hearing people, so the types of vowels and consonants in words were the target information.  

The second step is to decide \textbf{how} to convey the information. Since vibrations can be configured with different characteristics such as intensity, frequency, duration, position, and so on, the common practice is to break down the information and deliver it via multiple channels to maximize information transmission \cite{tan2020methodology}. For example, vowels and consonants for the phonemic displays were distinguished by using moving patterns and static patterns, while different consonants were distinguished by actuator position, frequency modulation and vibration duration. 

The third step is to decide \textbf{where} to install the system and derive the layout of vibration units. In cases where the information was spatially connected to the surrounding environment, the layout should have a good mapping between tactile positions and spatial direction. In the UAV teleoperation, locations of vibrations were perceptually mapped to the directions of incoming obstacle. Perceptual limits should also be considered, e.g., spatial acuity indicated the upper limits of layouts, as adding more actuators to the layout would not increase the information transmission. 

The fourth step is to \textbf{build} the system. Control and vibration units should be chosen based on the scale of the system (e.g., the small control unit for localized systems such as sleeves and headbands and the large control unit for long-term usage and body-scale systems such as immersive virtual experience and robot teleoperation). When designing vibration patterns, if the information involves complex waveforms, designers can utilize the Waveform Library in the GUI Editor to expedite the authoring process, similar to its use for the phonemic display. If the information only requires parameterized control, the Python server and APIs should be handy.

The final step is to \textbf{test} the system and \textbf{iterate} on the design. If the desired information is not transmitted as expected, designers can consider changing the layout or using different vibration characteristics to convey the information.

%% file: docs/9_conclusion.tex
\section{Conclusion}

The VibraForge toolkit enables scalability and expressivity via a modular design and a chain-connection method. Technical evaluation validated toolkit's design requirements, case studies demonstrated the design space, and usability study highlighted low technical barrier and customizability. The lessons learned from evaluations contributed to potential directions for toolkit refinement and generalized pipeline for creating spatialized vibrotactile systems. We are excited to see how researchers and designers will harness the toolkit to elevate their own research and push the boundaries of design innovation using spatialized vibrotactile feedback.

%% file: main.bbl

\begin{thebibliography}{64}

%%% the Web address.

\ifx \showCODEN    \undefined \def \showCODEN     #1{\unskip}     \fi
\ifx \showISBNx    \undefined \def \showISBNx     #1{\unskip}     \fi
\ifx \showISBNxiii \undefined \def \showISBNxiii  #1{\unskip}     \fi
\ifx \showISSN     \undefined \def \showISSN      #1{\unskip}     \fi
\ifx \showLCCN     \undefined \def \showLCCN      #1{\unskip}     \fi
\ifx \shownote     \undefined \def \shownote      #1{#1}          \fi
\ifx \showarticletitle \undefined \def \showarticletitle #1{#1}   \fi
\ifx \showURL      \undefined \def \showURL       {\relax}        \fi
% invisible to TeX
\providecommand\bibfield[2]{#2}
\providecommand\bibinfo[2]{#2}
\providecommand\natexlab[1]{#1}
\providecommand\showeprint[2][]{arXiv:#2}

\bibitem[Aggravi et~al\mbox{.}(2018)]%
        {aggravi2018design}
\bibfield{author}{\bibinfo{person}{Marco Aggravi}, \bibinfo{person}{Florent Pausé}, \bibinfo{person}{Paolo~Robuffo Giordano}, {and} \bibinfo{person}{Claudio Pacchierotti}.} \bibinfo{year}{2018}\natexlab{}.
\newblock \showarticletitle{Design and Evaluation of a Wearable Haptic Device for Skin Stretch, Pressure, and Vibrotactile Stimuli}.
\newblock \bibinfo{journal}{\emph{IEEE Robotics and Automation Letters}} \bibinfo{volume}{3}, \bibinfo{number}{3} (\bibinfo{year}{2018}), \bibinfo{pages}{2166--2173}.
\newblock
\href{https://doi.org/10.1109/LRA.2018.2810887}{doi:\nolinkurl{10.1109/LRA.2018.2810887}}


\bibitem[Ames et~al\mbox{.}(2019)]%
        {ames2019control}
\bibfield{author}{\bibinfo{person}{Aaron~D Ames}, \bibinfo{person}{Samuel Coogan}, \bibinfo{person}{Magnus Egerstedt}, \bibinfo{person}{Gennaro Notomista}, \bibinfo{person}{Koushil Sreenath}, {and} \bibinfo{person}{Paulo Tabuada}.} \bibinfo{year}{2019}\natexlab{}.
\newblock \showarticletitle{Control barrier functions: Theory and applications}. In \bibinfo{booktitle}{\emph{2019 18th European control conference (ECC)}}. IEEE, \bibinfo{pages}{3420--3431}.
\newblock


\bibitem[Bark et~al\mbox{.}(2014)]%
        {bark2014effects}
\bibfield{author}{\bibinfo{person}{Karlin Bark}, \bibinfo{person}{Emily Hyman}, \bibinfo{person}{Frank Tan}, \bibinfo{person}{Elizabeth Cha}, \bibinfo{person}{Steven~A Jax}, \bibinfo{person}{Laurel~J Buxbaum}, {and} \bibinfo{person}{Katherine~J Kuchenbecker}.} \bibinfo{year}{2014}\natexlab{}.
\newblock \showarticletitle{Effects of vibrotactile feedback on human learning of arm motions}.
\newblock \bibinfo{journal}{\emph{IEEE Transactions on Neural Systems and Rehabilitation Engineering}} \bibinfo{volume}{23}, \bibinfo{number}{1} (\bibinfo{year}{2014}), \bibinfo{pages}{51--63}.
\newblock


\bibitem[bHaptics(2024)]%
        {bHapticsWebsite}
\bibfield{author}{\bibinfo{person}{bHaptics}.} \bibinfo{year}{2024}\natexlab{}.
\newblock \bibinfo{title}{bHaptics: Tactile Feedback for VR, Gaming, and Music}.
\newblock
\urldef\tempurl%
\url{https://www.bhaptics.com/}
\showURL{%
\tempurl}


\bibitem[Brandt and Colton(2010)]%
        {brandt2009haptic}
\bibfield{author}{\bibinfo{person}{Adam~M. Brandt} {and} \bibinfo{person}{Mark~B. Colton}.} \bibinfo{year}{2010}\natexlab{}.
\newblock \showarticletitle{Haptic collision avoidance for a remotely operated quadrotor UAV in indoor environments}. In \bibinfo{booktitle}{\emph{2010 IEEE International Conference on Systems, Man and Cybernetics}}. \bibinfo{pages}{2724--2731}.
\newblock
\href{https://doi.org/10.1109/ICSMC.2010.5641798}{doi:\nolinkurl{10.1109/ICSMC.2010.5641798}}


\bibitem[Brooke(2013)]%
        {brooke2013sus}
\bibfield{author}{\bibinfo{person}{John Brooke}.} \bibinfo{year}{2013}\natexlab{}.
\newblock \showarticletitle{SUS: a retrospective.}
\newblock \bibinfo{journal}{\emph{Journal of usability studies}} \bibinfo{volume}{8}, \bibinfo{number}{2} (\bibinfo{year}{2013}).
\newblock


\bibitem[Choi and Kuchenbecker(2013)]%
        {choi2013vibrotactile}
\bibfield{author}{\bibinfo{person}{Seungmoon Choi} {and} \bibinfo{person}{Katherine~J. Kuchenbecker}.} \bibinfo{year}{2013}\natexlab{}.
\newblock \showarticletitle{Vibrotactile Display: Perception, Technology, and Applications}.
\newblock \bibinfo{journal}{\emph{Proc. IEEE}} \bibinfo{volume}{101}, \bibinfo{number}{9} (\bibinfo{year}{2013}), \bibinfo{pages}{2093--2104}.
\newblock
\href{https://doi.org/10.1109/JPROC.2012.2221071}{doi:\nolinkurl{10.1109/JPROC.2012.2221071}}


\bibitem[Dementyev et~al\mbox{.}(2021)]%
        {dementyev2021vhp}
\bibfield{author}{\bibinfo{person}{Artem Dementyev}, \bibinfo{person}{Pascal Getreuer}, \bibinfo{person}{Dimitri Kanevsky}, \bibinfo{person}{Malcolm Slaney}, {and} \bibinfo{person}{Richard~F Lyon}.} \bibinfo{year}{2021}\natexlab{}.
\newblock \showarticletitle{VHP: Vibrotactile Haptics Platform for On-body Applications}. In \bibinfo{booktitle}{\emph{The 34th Annual ACM Symposium on User Interface Software and Technology}} (Virtual Event, USA) \emph{(\bibinfo{series}{UIST '21})}. \bibinfo{publisher}{Association for Computing Machinery}, \bibinfo{address}{New York, NY, USA}, \bibinfo{pages}{598–612}.
\newblock
\showISBNx{9781450386357}
\href{https://doi.org/10.1145/3472749.3474772}{doi:\nolinkurl{10.1145/3472749.3474772}}


\bibitem[Elsayed et~al\mbox{.}(2020)]%
        {elsayed2020vibromap}
\bibfield{author}{\bibinfo{person}{Hesham Elsayed}, \bibinfo{person}{Martin Weigel}, \bibinfo{person}{Florian M{\"u}ller}, \bibinfo{person}{Martin Schmitz}, \bibinfo{person}{Karola Marky}, \bibinfo{person}{Sebastian G{\"u}nther}, \bibinfo{person}{Jan Riemann}, {and} \bibinfo{person}{Max M{\"u}hlh{\"a}user}.} \bibinfo{year}{2020}\natexlab{}.
\newblock \showarticletitle{Vibromap: Understanding the spacing of vibrotactile actuators across the body}.
\newblock \bibinfo{journal}{\emph{Proceedings of the ACM on Interactive, Mobile, Wearable and Ubiquitous Technologies}} \bibinfo{volume}{4}, \bibinfo{number}{4} (\bibinfo{year}{2020}), \bibinfo{pages}{1--16}.
\newblock


\bibitem[Fox(2008)]%
        {holeinthewall}
\bibfield{author}{\bibinfo{person}{Fox}.} \bibinfo{year}{2008}\natexlab{}.
\newblock \bibinfo{title}{Hole in the Wall (American game show)}.
\newblock
\urldef\tempurl%
\url{https://en.wikipedia.org/wiki/Hole_in_the_Wall_(American_game_show)}
\showURL{%
\tempurl}
\newblock
\shownote{Accessed: September, 2023}.


\bibitem[Hartmann et~al\mbox{.}(2006)]%
        {hartmann2006reflective}
\bibfield{author}{\bibinfo{person}{Bj{\"o}rn Hartmann}, \bibinfo{person}{Scott~R Klemmer}, \bibinfo{person}{Michael Bernstein}, \bibinfo{person}{Leith Abdulla}, \bibinfo{person}{Brandon Burr}, \bibinfo{person}{Avi Robinson-Mosher}, {and} \bibinfo{person}{Jennifer Gee}.} \bibinfo{year}{2006}\natexlab{}.
\newblock \showarticletitle{Reflective physical prototyping through integrated design, test, and analysis}. In \bibinfo{booktitle}{\emph{Proceedings of the 19th annual ACM symposium on User interface software and technology}}. \bibinfo{pages}{299--308}.
\newblock


\bibitem[Hou et~al\mbox{.}(2017)]%
        {hou2017dynamic}
\bibfield{author}{\bibinfo{person}{Xiaolei Hou}, \bibinfo{person}{Pengfei Fang}, {and} \bibinfo{person}{Yaohong Qu}.} \bibinfo{year}{2017}\natexlab{}.
\newblock \showarticletitle{Dynamic kinesthetic boundary for haptic teleoperation of unicycle type ground mobile robots}. In \bibinfo{booktitle}{\emph{2017 36th Cine. Control Conf. (CCC)}}. IEEE, \bibinfo{pages}{6246--6251}.
\newblock


\bibitem[Huang et~al\mbox{.}(2024a)]%
        {huang2024investigating}
\bibfield{author}{\bibinfo{person}{Bingjian Huang}, \bibinfo{person}{Paul~H. Dietz}, {and} \bibinfo{person}{Daniel Wigdor}.} \bibinfo{year}{2024}\natexlab{a}.
\newblock \showarticletitle{Investigating the Effects of Intensity and Frequency on Vibrotactile Spatial Acuity}.
\newblock \bibinfo{journal}{\emph{IEEE Transactions on Haptics}} (\bibinfo{year}{2024}), \bibinfo{pages}{1--13}.
\newblock
\href{https://doi.org/10.1109/TOH.2024.3350929}{doi:\nolinkurl{10.1109/TOH.2024.3350929}}


\bibitem[Huang et~al\mbox{.}(2024b)]%
        {huang2024aerohaptix}
\bibfield{author}{\bibinfo{person}{Bingjian Huang}, \bibinfo{person}{Zhecheng Wang}, \bibinfo{person}{Qilong Cheng}, \bibinfo{person}{Siyi Ren}, \bibinfo{person}{Hanfeng Cai}, \bibinfo{person}{Antonio~Alvarez Valdivia}, \bibinfo{person}{Karthik Mahadevan}, {and} \bibinfo{person}{Daniel Wigdor}.} \bibinfo{year}{2024}\natexlab{b}.
\newblock \showarticletitle{AeroHaptix: A Wearable Vibrotactile Feedback System for Enhancing Collision Avoidance in UAV Teleoperation}.
\newblock \bibinfo{journal}{\emph{arXiv preprint arXiv:2407.12105}} (\bibinfo{year}{2024}).
\newblock


\bibitem[Inc(2024)]%
        {novint_falcon}
\bibfield{author}{\bibinfo{person}{Novint~Technologies Inc}.} \bibinfo{year}{2024}\natexlab{}.
\newblock \bibinfo{title}{Novint Falcon}.
\newblock \bibinfo{howpublished}{\url{https://hapticshouse.com/pages/novints-falcon-haptic-device}}.
\newblock
\newblock
\shownote{Accessed: Feb 24, 2024}.


\bibitem[Israr et~al\mbox{.}(2016)]%
        {israr2016stereohaptics}
\bibfield{author}{\bibinfo{person}{Ali Israr}, \bibinfo{person}{Siyan Zhao}, \bibinfo{person}{Kyna McIntosh}, \bibinfo{person}{Zachary Schwemler}, \bibinfo{person}{Adam Fritz}, \bibinfo{person}{John Mars}, \bibinfo{person}{Job Bedford}, \bibinfo{person}{Christian Frisson}, \bibinfo{person}{Ivan Huerta}, \bibinfo{person}{Maggie Kosek}, \bibinfo{person}{Babis Koniaris}, {and} \bibinfo{person}{Kenny Mitchell}.} \bibinfo{year}{2016}\natexlab{}.
\newblock \showarticletitle{Stereohaptics: a haptic interaction toolkit for tangible virtual experiences}. In \bibinfo{booktitle}{\emph{ACM SIGGRAPH 2016 Studio}} (Anaheim, California) \emph{(\bibinfo{series}{SIGGRAPH '16})}. \bibinfo{publisher}{Association for Computing Machinery}, \bibinfo{address}{New York, NY, USA}, Article \bibinfo{articleno}{13}, \bibinfo{numpages}{57}~pages.
\newblock
\showISBNx{9781450343732}
\href{https://doi.org/10.1145/2929484.2970273}{doi:\nolinkurl{10.1145/2929484.2970273}}


\bibitem[Jo et~al\mbox{.}(2023)]%
        {jo2023trainertap}
\bibfield{author}{\bibinfo{person}{Hye-Young Jo}, \bibinfo{person}{Chan~Hu Wie}, \bibinfo{person}{Yejin Jang}, \bibinfo{person}{Dong-Uk Kim}, \bibinfo{person}{Yurim Son}, {and} \bibinfo{person}{Yoonji Kim}.} \bibinfo{year}{2023}\natexlab{}.
\newblock \showarticletitle{TrainerTap: Weightlifting Support System Prototype Simulating Personal Trainer's Tactile and Auditory Guidance}. In \bibinfo{booktitle}{\emph{Adjunct Proceedings of the 36th Annual ACM Symposium on User Interface Software and Technology}}. \bibinfo{pages}{1--3}.
\newblock


\bibitem[Jung et~al\mbox{.}(2022)]%
        {jung2022wireless}
\bibfield{author}{\bibinfo{person}{Yei~Hwan Jung}, \bibinfo{person}{Jae-Young Yoo}, \bibinfo{person}{Abraham V{\'a}zquez-Guardado}, \bibinfo{person}{Jae-Hwan Kim}, \bibinfo{person}{Jin-Tae Kim}, \bibinfo{person}{Haiwen Luan}, \bibinfo{person}{Minsu Park}, \bibinfo{person}{Jaeman Lim}, \bibinfo{person}{Hee-Sup Shin}, \bibinfo{person}{Chun-Ju Su}, {et~al\mbox{.}}} \bibinfo{year}{2022}\natexlab{}.
\newblock \showarticletitle{A wireless haptic interface for programmable patterns of touch across large areas of the skin}.
\newblock \bibinfo{journal}{\emph{Nature Electronics}} \bibinfo{volume}{5}, \bibinfo{number}{6} (\bibinfo{year}{2022}), \bibinfo{pages}{374--385}.
\newblock


\bibitem[Kaul and Rohs(2017)]%
        {kaul2017haptichead}
\bibfield{author}{\bibinfo{person}{Oliver~Beren Kaul} {and} \bibinfo{person}{Michael Rohs}.} \bibinfo{year}{2017}\natexlab{}.
\newblock \showarticletitle{Haptichead: A spherical vibrotactile grid around the head for 3d guidance in virtual and augmented reality}. In \bibinfo{booktitle}{\emph{Proceedings of the 2017 CHI Conference on Human Factors in Computing Systems}}. \bibinfo{pages}{3729--3740}.
\newblock


\bibitem[Lab(2019)]%
        {ohshape}
\bibfield{author}{\bibinfo{person}{Odders Lab}.} \bibinfo{year}{2019}\natexlab{}.
\newblock \bibinfo{title}{OhShape}.
\newblock
\urldef\tempurl%
\url{https://store.steampowered.com/app/1098100/OhShape/}
\showURL{%
\tempurl}
\newblock
\shownote{Accessed: September, 2023}.


\bibitem[Ledo et~al\mbox{.}(2018)]%
        {ledo2018evaluation}
\bibfield{author}{\bibinfo{person}{David Ledo}, \bibinfo{person}{Steven Houben}, \bibinfo{person}{Jo Vermeulen}, \bibinfo{person}{Nicolai Marquardt}, \bibinfo{person}{Lora Oehlberg}, {and} \bibinfo{person}{Saul Greenberg}.} \bibinfo{year}{2018}\natexlab{}.
\newblock \showarticletitle{Evaluation Strategies for HCI Toolkit Research}. In \bibinfo{booktitle}{\emph{Proceedings of the 2018 CHI Conference on Human Factors in Computing Systems}} (Montreal QC, Canada) \emph{(\bibinfo{series}{CHI '18})}. \bibinfo{publisher}{Association for Computing Machinery}, \bibinfo{address}{New York, NY, USA}, \bibinfo{pages}{1–17}.
\newblock
\showISBNx{9781450356206}
\href{https://doi.org/10.1145/3173574.3173610}{doi:\nolinkurl{10.1145/3173574.3173610}}


\bibitem[Lemmens et~al\mbox{.}(2009)]%
        {lemmens2009body}
\bibfield{author}{\bibinfo{person}{Paul Lemmens}, \bibinfo{person}{Floris Crompvoets}, \bibinfo{person}{Dirk Brokken}, \bibinfo{person}{Jack Van Den~Eerenbeemd}, {and} \bibinfo{person}{Gert-Jan de Vries}.} \bibinfo{year}{2009}\natexlab{}.
\newblock \showarticletitle{A body-conforming tactile jacket to enrich movie viewing}. In \bibinfo{booktitle}{\emph{World Haptics 2009-Third Joint EuroHaptics conference and Symposium on Haptic Interfaces for Virtual Environment and Teleoperator Systems}}. IEEE, \bibinfo{pages}{7--12}.
\newblock


\bibitem[LI et~al\mbox{.}(2013)]%
        {li2013design}
\bibfield{author}{\bibinfo{person}{Yan LI}, \bibinfo{person}{Yuki OBATA}, \bibinfo{person}{Miyuki KUMAGAI}, \bibinfo{person}{Marina ISHIKAWA}, \bibinfo{person}{Moeki OWAKI}, \bibinfo{person}{Natsuki FUKAMI}, {and} \bibinfo{person}{Kiyoshi TOMIMATSU}.} \bibinfo{year}{2013}\natexlab{}.
\newblock \showarticletitle{A design study for the haptic vest as a navigation system}.
\newblock \bibinfo{journal}{\emph{International Journal of Asia Digital Art and Design Association}} \bibinfo{volume}{17}, \bibinfo{number}{1} (\bibinfo{year}{2013}), \bibinfo{pages}{10--17}.
\newblock


\bibitem[Lindeman et~al\mbox{.}(2004)]%
        {lindeman2004towards}
\bibfield{author}{\bibinfo{person}{Robert~W. Lindeman}, \bibinfo{person}{Robert Page}, \bibinfo{person}{Yasuyuki Yanagida}, {and} \bibinfo{person}{John~L. Sibert}.} \bibinfo{year}{2004}\natexlab{}.
\newblock \showarticletitle{Towards full-body haptic feedback: the design and deployment of a spatialized vibrotactile feedback system}. In \bibinfo{booktitle}{\emph{Proceedings of the ACM Symposium on Virtual Reality Software and Technology}} (Hong Kong) \emph{(\bibinfo{series}{VRST '04})}. \bibinfo{publisher}{Association for Computing Machinery}, \bibinfo{address}{New York, NY, USA}, \bibinfo{pages}{146–149}.
\newblock
\showISBNx{1581139071}
\href{https://doi.org/10.1145/1077534.1077562}{doi:\nolinkurl{10.1145/1077534.1077562}}


\bibitem[Lindeman et~al\mbox{.}(2006)]%
        {lindeman2006tactapack}
\bibfield{author}{\bibinfo{person}{Robert~W Lindeman}, \bibinfo{person}{Yasuyuki Yanagida}, \bibinfo{person}{Kenichi Hosaka}, {and} \bibinfo{person}{Shinji Abe}.} \bibinfo{year}{2006}\natexlab{}.
\newblock \showarticletitle{The TactaPack: A wireless sensor/actuator package for physical therapy applications}. In \bibinfo{booktitle}{\emph{2006 14th Symposium on Haptic Interfaces for Virtual Environment and Teleoperator Systems}}. IEEE, \bibinfo{pages}{337--341}.
\newblock


\bibitem[Louison et~al\mbox{.}(2017)]%
        {louison2017spatialized}
\bibfield{author}{\bibinfo{person}{Cephise Louison}, \bibinfo{person}{Fabien Ferlay}, {and} \bibinfo{person}{Daniel~R Mestre}.} \bibinfo{year}{2017}\natexlab{}.
\newblock \showarticletitle{Spatialized vibrotactile feedback contributes to goal-directed movements in cluttered virtual environments}. In \bibinfo{booktitle}{\emph{2017 IEEE Symp. on 3D User Interfaces (3DUI)}}. IEEE, \bibinfo{pages}{99--102}.
\newblock


\bibitem[Machinery and Co.(2022)]%
        {JinlongWebsite}
\bibfield{author}{\bibinfo{person}{Jinlong Machinery} {and} \bibinfo{person}{Electronics. Co.}} \bibinfo{year}{2022}\natexlab{}.
\newblock \bibinfo{title}{LRA - Haptic Motors}.
\newblock
\urldef\tempurl%
\url{http://en.kotl.com.cn/prod_view.aspx?TypeId=75&Id=252&Fid=t3:75:3}
\showURL{%
\tempurl}


\bibitem[Mart{\'\i}nez et~al\mbox{.}(2014)]%
        {martinez2014vitaki}
\bibfield{author}{\bibinfo{person}{Jonatan Mart{\'\i}nez}, \bibinfo{person}{Arturo~S Garc{\'\i}a}, \bibinfo{person}{Miguel Oliver}, \bibinfo{person}{Jos{\'e}~P Molina}, {and} \bibinfo{person}{Pascual Gonz{\'a}lez}.} \bibinfo{year}{2014}\natexlab{}.
\newblock \showarticletitle{Vitaki: a vibrotactile prototyping toolkit for virtual reality and video games}.
\newblock \bibinfo{journal}{\emph{International Journal of Human-Computer Interaction}} \bibinfo{volume}{30}, \bibinfo{number}{11} (\bibinfo{year}{2014}), \bibinfo{pages}{855--871}.
\newblock


\bibitem[Meta~Platforms(2023a)]%
        {oculus2023haptics}
\bibfield{author}{\bibinfo{person}{Inc. Meta~Platforms}.} \bibinfo{year}{2023}\natexlab{a}.
\newblock \bibinfo{title}{Haptics SDK Studio for Meta Quest VR}.
\newblock
\urldef\tempurl%
\url{https://developer.oculus.com/blog/haptics-sdk-studio-meta-quest-vr/}
\showURL{%
\tempurl}
\newblock
\shownote{Accessed: 2024-08-26}.


\bibitem[Meta~Platforms(2023b)]%
        {MetaQuest2}
\bibfield{author}{\bibinfo{person}{Inc Meta~Platforms}.} \bibinfo{year}{2023}\natexlab{b}.
\newblock \bibinfo{title}{Meta Quest 2}.
\newblock
\urldef\tempurl%
\url{https://www.meta.com/ca/quest/products/quest-2/?utm_source=www.google.com&utm_medium=oculusredirect}
\showURL{%
\tempurl}


\bibitem[Minamizawa et~al\mbox{.}(2012)]%
        {minamizawa2012techtile}
\bibfield{author}{\bibinfo{person}{Kouta Minamizawa}, \bibinfo{person}{Yasuaki Kakehi}, \bibinfo{person}{Masashi Nakatani}, \bibinfo{person}{Soichiro Mihara}, {and} \bibinfo{person}{Susumu Tachi}.} \bibinfo{year}{2012}\natexlab{}.
\newblock \showarticletitle{TECHTILE toolkit: a prototyping tool for design and education of haptic media}. In \bibinfo{booktitle}{\emph{Proceedings of the 2012 Virtual Reality International Conference}}. \bibinfo{pages}{1--2}.
\newblock


\bibitem[Myers et~al\mbox{.}(2000)]%
        {myers2000past}
\bibfield{author}{\bibinfo{person}{Brad Myers}, \bibinfo{person}{Scott~E Hudson}, {and} \bibinfo{person}{Randy Pausch}.} \bibinfo{year}{2000}\natexlab{}.
\newblock \showarticletitle{Past, present, and future of user interface software tools}.
\newblock \bibinfo{journal}{\emph{ACM Transactions on Computer-Human Interaction (TOCHI)}} \bibinfo{volume}{7}, \bibinfo{number}{1} (\bibinfo{year}{2000}), \bibinfo{pages}{3--28}.
\newblock


\bibitem[Omari et~al\mbox{.}(2013)]%
        {omari2013bilateral}
\bibfield{author}{\bibinfo{person}{Sammy Omari}, \bibinfo{person}{Minh-Duc Hua}, \bibinfo{person}{Guillaume Ducard}, {and} \bibinfo{person}{Tarek Hamel}.} \bibinfo{year}{2013}\natexlab{}.
\newblock \showarticletitle{Bilateral haptic teleoperation of VTOL UAVs}. In \bibinfo{booktitle}{\emph{2013 IEEE Int. Conf. on Robot. and Automat.}} IEEE, \bibinfo{pages}{2393--2399}.
\newblock


\bibitem[Pacchierotti and Prattichizzo(2024)]%
        {pacchierotti2024cutaneous}
\bibfield{author}{\bibinfo{person}{Claudio Pacchierotti} {and} \bibinfo{person}{Domenico Prattichizzo}.} \bibinfo{year}{2024}\natexlab{}.
\newblock \showarticletitle{Cutaneous/Tactile Haptic Feedback in Robotic Teleoperation: Motivation, Survey, and Perspectives}.
\newblock \bibinfo{journal}{\emph{IEEE Transactions on Robotics}}  \bibinfo{volume}{40} (\bibinfo{year}{2024}), \bibinfo{pages}{978--998}.
\newblock
\href{https://doi.org/10.1109/TRO.2023.3344027}{doi:\nolinkurl{10.1109/TRO.2023.3344027}}


\bibitem[Park et~al\mbox{.}(2023)]%
        {park2023information}
\bibfield{author}{\bibinfo{person}{Jaejun Park}, \bibinfo{person}{Junwoo Kim}, \bibinfo{person}{Sangyoon Han}, \bibinfo{person}{Chaeyong Park}, \bibinfo{person}{Junseok Park}, {and} \bibinfo{person}{Seungmoon Choi}.} \bibinfo{year}{2023}\natexlab{}.
\newblock \showarticletitle{Information Transfer of Full-Body Vibrotactile Stimuli: An Initial Study with One to Three Sequential Vibrations}. In \bibinfo{booktitle}{\emph{2023 IEEE World Haptics Conference (WHC)}}. IEEE, \bibinfo{pages}{41--47}.
\newblock


\bibitem[Pezent et~al\mbox{.}(2020)]%
        {pezent2020syntacts}
\bibfield{author}{\bibinfo{person}{Evan Pezent}, \bibinfo{person}{Brandon Cambio}, {and} \bibinfo{person}{Marcia~K O’Malley}.} \bibinfo{year}{2020}\natexlab{}.
\newblock \showarticletitle{Syntacts: Open-source software and hardware for audio-controlled haptics}.
\newblock \bibinfo{journal}{\emph{IEEE Transactions on Haptics}} \bibinfo{volume}{14}, \bibinfo{number}{1} (\bibinfo{year}{2020}), \bibinfo{pages}{225--233}.
\newblock


\bibitem[Pongrac(2006)]%
        {pongrac2006vibrotactile}
\bibfield{author}{\bibinfo{person}{Helena Pongrac}.} \bibinfo{year}{2006}\natexlab{}.
\newblock \showarticletitle{Vibrotactile perception: Differential effects of frequency, amplitude, and acceleration}. In \bibinfo{booktitle}{\emph{2006 ieee international workshop on haptic audio visual environments and their applications (have 2006)}}. IEEE, \bibinfo{pages}{54--59}.
\newblock


\bibitem[Prabhu et~al\mbox{.}(2020)]%
        {prabhu2020vibrosleeve}
\bibfield{author}{\bibinfo{person}{Deepa Prabhu}, \bibinfo{person}{Muhammad~Mehedi Hasan}, \bibinfo{person}{Lisa Wise}, \bibinfo{person}{Clare MacMahon}, {and} \bibinfo{person}{Chris McCarthy}.} \bibinfo{year}{2020}\natexlab{}.
\newblock \showarticletitle{VibroSleeve: A wearable vibro-tactile feedback device for arm guidance}. In \bibinfo{booktitle}{\emph{2020 42nd Annual International Conference of the IEEE Engineering in Medicine \& Biology Society (EMBC)}}. IEEE, \bibinfo{pages}{4909--4912}.
\newblock


\bibitem[PUI~Audio(2023)]%
        {PUIAudio}
\bibfield{author}{\bibinfo{person}{Inc. PUI~Audio}.} \bibinfo{year}{2023}\natexlab{}.
\newblock \bibinfo{title}{HD-VA3222}.
\newblock
\urldef\tempurl%
\url{https://puiaudio.com/product/haptics/hd-va3222}
\showURL{%
\tempurl}
\newblock
\shownote{Accessed: September, 2023}.


\bibitem[Reed et~al\mbox{.}(2018)]%
        {reed2018phonemic}
\bibfield{author}{\bibinfo{person}{Charlotte~M Reed}, \bibinfo{person}{Hong~Z Tan}, \bibinfo{person}{Zachary~D Perez}, \bibinfo{person}{E~Courtenay Wilson}, \bibinfo{person}{Frederico~M Severgnini}, \bibinfo{person}{Jaehong Jung}, \bibinfo{person}{Juan~S Martinez}, \bibinfo{person}{Yang Jiao}, \bibinfo{person}{Ali Israr}, \bibinfo{person}{Frances Lau}, {et~al\mbox{.}}} \bibinfo{year}{2018}\natexlab{}.
\newblock \showarticletitle{A phonemic-based tactile display for speech communication}.
\newblock \bibinfo{journal}{\emph{IEEE transactions on haptics}} \bibinfo{volume}{12}, \bibinfo{number}{1} (\bibinfo{year}{2018}), \bibinfo{pages}{2--17}.
\newblock


\bibitem[Reyes et~al\mbox{.}(2015)]%
        {reyes2015outdoor}
\bibfield{author}{\bibinfo{person}{Sergio Reyes}, \bibinfo{person}{Hugo Romero}, \bibinfo{person}{Sergio Salazar}, \bibinfo{person}{Rogelio Lozano}, {and} \bibinfo{person}{Omar Santos}.} \bibinfo{year}{2015}\natexlab{}.
\newblock \showarticletitle{Outdoor haptic teleoperation of a hexarotor UAV}. In \bibinfo{booktitle}{\emph{2015 Int. Conf. on Unmanned Aircr. Syst. (ICUAS)}}. IEEE, \bibinfo{pages}{972--979}.
\newblock


\bibitem[Schneider et~al\mbox{.}(2015)]%
        {schneider2015tactile}
\bibfield{author}{\bibinfo{person}{Oliver~S. Schneider}, \bibinfo{person}{Ali Israr}, {and} \bibinfo{person}{Karon~E. MacLean}.} \bibinfo{year}{2015}\natexlab{}.
\newblock \showarticletitle{Tactile Animation by Direct Manipulation of Grid Displays}. In \bibinfo{booktitle}{\emph{Proceedings of the 28th Annual ACM Symposium on User Interface Software \& Technology}} (Charlotte, NC, USA) \emph{(\bibinfo{series}{UIST '15})}. \bibinfo{publisher}{Association for Computing Machinery}, \bibinfo{address}{New York, NY, USA}, \bibinfo{pages}{21–30}.
\newblock
\showISBNx{9781450337793}
\href{https://doi.org/10.1145/2807442.2807470}{doi:\nolinkurl{10.1145/2807442.2807470}}


\bibitem[Schneider and MacLean(2016)]%
        {schneider2016studying}
\bibfield{author}{\bibinfo{person}{Oliver~S Schneider} {and} \bibinfo{person}{Karon~E MacLean}.} \bibinfo{year}{2016}\natexlab{}.
\newblock \showarticletitle{Studying design process and example use with Macaron, a web-based vibrotactile effect editor}. In \bibinfo{booktitle}{\emph{2016 IEEE Haptics Symposium (Haptics)}}. IEEE, \bibinfo{pages}{52--58}.
\newblock


\bibitem[Seifi et~al\mbox{.}(2015)]%
        {seifi2015vibviz}
\bibfield{author}{\bibinfo{person}{Hasti Seifi}, \bibinfo{person}{Kailun Zhang}, {and} \bibinfo{person}{Karon~E. MacLean}.} \bibinfo{year}{2015}\natexlab{}.
\newblock \showarticletitle{VibViz: Organizing, visualizing and navigating vibration libraries}. In \bibinfo{booktitle}{\emph{2015 IEEE World Haptics Conference (WHC)}}. \bibinfo{pages}{254--259}.
\newblock
\href{https://doi.org/10.1109/WHC.2015.7177722}{doi:\nolinkurl{10.1109/WHC.2015.7177722}}


\bibitem[Shah et~al\mbox{.}(2018)]%
        {airsim2017fsr}
\bibfield{author}{\bibinfo{person}{Shital Shah}, \bibinfo{person}{Debadeepta Dey}, \bibinfo{person}{Chris Lovett}, {and} \bibinfo{person}{Ashish Kapoor}.} \bibinfo{year}{2018}\natexlab{}.
\newblock \showarticletitle{AirSim: High-Fidelity Visual and Physical Simulation for Autonomous Vehicles}. In \bibinfo{booktitle}{\emph{Field and Service Robotics}}, \bibfield{editor}{\bibinfo{person}{Marco Hutter} {and} \bibinfo{person}{Roland Siegwart}} (Eds.). \bibinfo{publisher}{Springer International Publishing}, \bibinfo{address}{Cham}, \bibinfo{pages}{621--635}.
\newblock
\showISBNx{978-3-319-67361-5}


\bibitem[Shtarbanov et~al\mbox{.}(2023)]%
        {shtarbanov2023sleeveio}
\bibfield{author}{\bibinfo{person}{Ali Shtarbanov}, \bibinfo{person}{Mengjia Zhu}, \bibinfo{person}{Nicholas Colonnese}, {and} \bibinfo{person}{Amirhossein Hajiagha~Memar}.} \bibinfo{year}{2023}\natexlab{}.
\newblock \showarticletitle{SleeveIO: Modular and Reconfigurable Platform for Multimodal Wearable Haptic Feedback Interactions}. In \bibinfo{booktitle}{\emph{Proceedings of the 36th Annual ACM Symposium on User Interface Software and Technology}} (San Francisco, CA, USA) \emph{(\bibinfo{series}{UIST '23})}. \bibinfo{publisher}{Association for Computing Machinery}, \bibinfo{address}{New York, NY, USA}, Article \bibinfo{articleno}{70}, \bibinfo{numpages}{15}~pages.
\newblock
\showISBNx{9798400701320}
\href{https://doi.org/10.1145/3586183.3606739}{doi:\nolinkurl{10.1145/3586183.3606739}}


\bibitem[SNAME(2022)]%
        {xia2022virtual}
SNAME \bibinfo{year}{2022}\natexlab{}.
\newblock \bibinfo{booktitle}{\emph{Virtual Telepresence for the Future of ROV Teleoperations: Opportunities and Challenges}}. \bibinfo{series}{SNAME Offshore Symposium}, Vol.~\bibinfo{volume}{Day 1 Tue, February 22, 2022}. SNAME.
\newblock
\href{https://doi.org/10.5957/TOS-2022-015}{doi:\nolinkurl{10.5957/TOS-2022-015}}
\showeprint{https://onepetro.org/SNAMETOS/proceedings-pdf/TOS22/1-TOS22/D011S001R001/2726486/sname-tos-2022-015.pdf}


\bibitem[{Supernatural}(2024)]%
        {supernaturalvr}
\bibfield{author}{\bibinfo{person}{{Supernatural}}.} \bibinfo{year}{2024}\natexlab{}.
\newblock \bibinfo{title}{Supernatural - Virtual Reality Fitness}.
\newblock \bibinfo{howpublished}{\url{https://www.getsupernatural.com/}}.
\newblock
\newblock
\shownote{Accessed: 2024-09-02}.


\bibitem[Sykownik and Masuch(2020)]%
        {sykownik2020experience}
\bibfield{author}{\bibinfo{person}{Philipp Sykownik} {and} \bibinfo{person}{Maic Masuch}.} \bibinfo{year}{2020}\natexlab{}.
\newblock \showarticletitle{The Experience of Social Touch in Multi-User Virtual Reality}. In \bibinfo{booktitle}{\emph{Proceedings of the 26th ACM Symposium on Virtual Reality Software and Technology}} (Virtual Event, Canada) \emph{(\bibinfo{series}{VRST '20})}. \bibinfo{publisher}{Association for Computing Machinery}, \bibinfo{address}{New York, NY, USA}, Article \bibinfo{articleno}{30}, \bibinfo{numpages}{11}~pages.
\newblock
\showISBNx{9781450376198}
\href{https://doi.org/10.1145/3385956.3418944}{doi:\nolinkurl{10.1145/3385956.3418944}}


\bibitem[Sánchez and Bohne(2015)]%
        {OpenVNAVI}
\bibfield{author}{\bibinfo{person}{David~Antón Sánchez} {and} \bibinfo{person}{René Bohne}.} \bibinfo{year}{2015}\natexlab{}.
\newblock \bibinfo{title}{OpenVNAVI: A Vibrotactile Navigation Aid for the Visually Impaired}.
\newblock
\urldef\tempurl%
\url{https://hci.rwth-aachen.de/openvnavi}
\showURL{%
\tempurl}
\newblock
\shownote{Accessed: September, 2023}.


\bibitem[Tan et~al\mbox{.}(2020)]%
        {tan2020methodology}
\bibfield{author}{\bibinfo{person}{Hong~Z Tan}, \bibinfo{person}{Seungmoon Choi}, \bibinfo{person}{Frances~WY Lau}, {and} \bibinfo{person}{Freddy Abnousi}.} \bibinfo{year}{2020}\natexlab{}.
\newblock \showarticletitle{Methodology for maximizing information transmission of haptic devices: A survey}.
\newblock \bibinfo{journal}{\emph{Proc. IEEE}} \bibinfo{volume}{108}, \bibinfo{number}{6} (\bibinfo{year}{2020}), \bibinfo{pages}{945--965}.
\newblock


\bibitem[Technologies(2020)]%
        {Unity2020_3_31}
\bibfield{author}{\bibinfo{person}{Unity Technologies}.} \bibinfo{year}{2020}\natexlab{}.
\newblock \bibinfo{title}{Unity Editor: What's New in 2020.3.31}.
\newblock
\urldef\tempurl%
\url{https://unity.com/releases/editor/whats-new/2020.3.31}
\showURL{%
\tempurl}


\bibitem[Terenti and Vatavu(2023)]%
        {terenti2023vireo}
\bibfield{author}{\bibinfo{person}{Mihail Terenti} {and} \bibinfo{person}{Radu-Daniel Vatavu}.} \bibinfo{year}{2023}\natexlab{}.
\newblock \showarticletitle{VIREO: Web-based Graphical Authoring of Vibrotactile Feedback for Interactions with Mobile and Wearable Devices}.
\newblock \bibinfo{journal}{\emph{International Journal of Human--Computer Interaction}} \bibinfo{volume}{39}, \bibinfo{number}{20} (\bibinfo{year}{2023}), \bibinfo{pages}{4162--4180}.
\newblock


\bibitem[Verrillo et~al\mbox{.}(1969)]%
        {verrillo1969sensation}
\bibfield{author}{\bibinfo{person}{Ronald~T Verrillo}, \bibinfo{person}{Anthony~J Fraioli}, {and} \bibinfo{person}{Robert~L Smith}.} \bibinfo{year}{1969}\natexlab{}.
\newblock \showarticletitle{Sensation magnitude of vibrotactile stimuli}.
\newblock \bibinfo{journal}{\emph{Perception \& Psychophysics}} \bibinfo{volume}{6}, \bibinfo{number}{6} (\bibinfo{year}{1969}), \bibinfo{pages}{366--372}.
\newblock


\bibitem[Vicon(2023)]%
        {Vicon}
\bibfield{author}{\bibinfo{person}{Vicon}.} \bibinfo{year}{2023}\natexlab{}.
\newblock \bibinfo{title}{Vicon Shogun Software System}.
\newblock
\urldef\tempurl%
\url{https://www.vicon.com/software/shogun/}
\showURL{%
\tempurl}
\newblock
\shownote{Accessed: September, 2023}.


\bibitem[Vogels(2004)]%
        {vogels2004detection}
\bibfield{author}{\bibinfo{person}{Ingrid~MLC Vogels}.} \bibinfo{year}{2004}\natexlab{}.
\newblock \showarticletitle{Detection of temporal delays in visual-haptic interfaces}.
\newblock \bibinfo{journal}{\emph{Human Factors}} \bibinfo{volume}{46}, \bibinfo{number}{1} (\bibinfo{year}{2004}), \bibinfo{pages}{118--134}.
\newblock


\bibitem[Wang et~al\mbox{.}(2021)]%
        {Wang2021Research}
\bibfield{author}{\bibinfo{person}{Ke Wang}, \bibinfo{person}{Yi-Hsuan Li}, \bibinfo{person}{Chun-Chen Hsu}, \bibinfo{person}{Jiabei Jiang}, \bibinfo{person}{Yan Liu}, \bibinfo{person}{Zirui Zhao}, \bibinfo{person}{Wei Yue}, {and} \bibinfo{person}{Lu Yao}.} \bibinfo{year}{2021}\natexlab{}.
\newblock \showarticletitle{A Research on Sensing Localization and Orientation of Objects in VR with Facial Vibrotactile Display}. In \bibinfo{booktitle}{\emph{Virtual, Augmented and Mixed Reality}}, \bibfield{editor}{\bibinfo{person}{Jessie Y.~C. Chen} {and} \bibinfo{person}{Gino Fragomeni}} (Eds.). \bibinfo{publisher}{Springer International Publishing}, \bibinfo{address}{Cham}, \bibinfo{pages}{223--241}.
\newblock
\showISBNx{978-3-030-77599-5}


\bibitem[Weber(1996)]%
        {weber1996eh}
\bibfield{author}{\bibinfo{person}{Ernst~Heinrich Weber}.} \bibinfo{year}{1996}\natexlab{}.
\newblock \bibinfo{booktitle}{\emph{EH Weber on the tactile senses}}.
\newblock \bibinfo{publisher}{Psychology Press}.
\newblock


\bibitem[West et~al\mbox{.}(2019)]%
        {west2019design}
\bibfield{author}{\bibinfo{person}{Travis~J. West}, \bibinfo{person}{Alexandra Bachmayer}, \bibinfo{person}{Sandeep Bhagwati}, \bibinfo{person}{Joanna Berzowska}, {and} \bibinfo{person}{Marcelo~M. Wanderley}.} \bibinfo{year}{2019}\natexlab{}.
\newblock \showarticletitle{The Design of the Body:Suit:Score, a Full-Body Vibrotactile Musical Score}. In \bibinfo{booktitle}{\emph{Human Interface and the Management of Information. Information in Intelligent Systems}}, \bibfield{editor}{\bibinfo{person}{Sakae Yamamoto} {and} \bibinfo{person}{Hirohiko Mori}} (Eds.). \bibinfo{publisher}{Springer International Publishing}, \bibinfo{address}{Cham}, \bibinfo{pages}{70--89}.
\newblock
\showISBNx{978-3-030-22649-7}


\bibitem[Wittchen et~al\mbox{.}(2022)]%
        {wittchen2022tactjam}
\bibfield{author}{\bibinfo{person}{Dennis Wittchen}, \bibinfo{person}{Katta Spiel}, \bibinfo{person}{Bruno Fruchard}, \bibinfo{person}{Donald Degraen}, \bibinfo{person}{Oliver Schneider}, \bibinfo{person}{Georg Freitag}, {and} \bibinfo{person}{Paul Strohmeier}.} \bibinfo{year}{2022}\natexlab{}.
\newblock \showarticletitle{TactJam: An End-to-End Prototyping Suite for Collaborative Design of On-Body Vibrotactile Feedback}. In \bibinfo{booktitle}{\emph{Proceedings of the Sixteenth International Conference on Tangible, Embedded, and Embodied Interaction}} (Daejeon, Republic of Korea) \emph{(\bibinfo{series}{TEI '22})}. \bibinfo{publisher}{Association for Computing Machinery}, \bibinfo{address}{New York, NY, USA}, Article \bibinfo{articleno}{1}, \bibinfo{numpages}{13}~pages.
\newblock
\showISBNx{9781450391474}
\href{https://doi.org/10.1145/3490149.3501307}{doi:\nolinkurl{10.1145/3490149.3501307}}


\bibitem[Witteveen et~al\mbox{.}(2015)]%
        {witteveen2015vibrotactile}
\bibfield{author}{\bibinfo{person}{Heidi~JB Witteveen}, \bibinfo{person}{Hans~S Rietman}, {and} \bibinfo{person}{Peter~H Veltink}.} \bibinfo{year}{2015}\natexlab{}.
\newblock \showarticletitle{Vibrotactile grasping force and hand aperture feedback for myoelectric forearm prosthesis users}.
\newblock \bibinfo{journal}{\emph{Prosthetics and orthotics international}} \bibinfo{volume}{39}, \bibinfo{number}{3} (\bibinfo{year}{2015}), \bibinfo{pages}{204--212}.
\newblock


\bibitem[Wu et~al\mbox{.}(2023)]%
        {wu2023ar}
\bibfield{author}{\bibinfo{person}{Yihong Wu}, \bibinfo{person}{Lingyun Yu}, \bibinfo{person}{Jie Xu}, \bibinfo{person}{Dazhen Deng}, \bibinfo{person}{Jiachen Wang}, \bibinfo{person}{Xiao Xie}, \bibinfo{person}{Hui Zhang}, {and} \bibinfo{person}{Yingcai Wu}.} \bibinfo{year}{2023}\natexlab{}.
\newblock \showarticletitle{AR-Enhanced Workouts: Exploring Visual Cues for At-Home Workout Videos in AR Environment}. In \bibinfo{booktitle}{\emph{Proceedings of the 36th Annual ACM Symposium on User Interface Software and Technology}} (San Francisco, CA, USA) \emph{(\bibinfo{series}{UIST '23})}. \bibinfo{publisher}{Association for Computing Machinery}, \bibinfo{address}{New York, NY, USA}, Article \bibinfo{articleno}{121}, \bibinfo{numpages}{15}~pages.
\newblock
\showISBNx{9798400701320}
\href{https://doi.org/10.1145/3586183.3606796}{doi:\nolinkurl{10.1145/3586183.3606796}}


\bibitem[Zhang et~al\mbox{.}(2020)]%
        {zhang2020haptic}
\bibfield{author}{\bibinfo{person}{Dawei Zhang}, \bibinfo{person}{Guang Yang}, {and} \bibinfo{person}{Rebecca~P Khurshid}.} \bibinfo{year}{2020}\natexlab{}.
\newblock \showarticletitle{Haptic teleoperation of uavs through control barrier functions}.
\newblock \bibinfo{journal}{\emph{IEEE Trans. on Haptics}} \bibinfo{volume}{13}, \bibinfo{number}{1} (\bibinfo{year}{2020}), \bibinfo{pages}{109--115}.
\newblock


\bibitem[Zhou et~al\mbox{.}(2023)]%
        {zhou2023tactorbots}
\bibfield{author}{\bibinfo{person}{Ran Zhou}, \bibinfo{person}{Zachary Schwemler}, \bibinfo{person}{Akshay Baweja}, \bibinfo{person}{Harpreet Sareen}, \bibinfo{person}{Casey~Lee Hunt}, {and} \bibinfo{person}{Daniel Leithinger}.} \bibinfo{year}{2023}\natexlab{}.
\newblock \showarticletitle{TactorBots: A Haptic Design Toolkit for Out-of-lab Exploration of Emotional Robotic Touch}. In \bibinfo{booktitle}{\emph{Proceedings of the 2023 CHI Conference on Human Factors in Computing Systems}} (Hamburg, Germany) \emph{(\bibinfo{series}{CHI '23})}. \bibinfo{publisher}{Association for Computing Machinery}, \bibinfo{address}{New York, NY, USA}, Article \bibinfo{articleno}{370}, \bibinfo{numpages}{19}~pages.
\newblock
\showISBNx{9781450394215}
\href{https://doi.org/10.1145/3544548.3580799}{doi:\nolinkurl{10.1145/3544548.3580799}}


\end{thebibliography}
